\documentclass
[twocolumn,aps,prl,preprintnumbers,amsmath,amssymb]{revtex4}

\usepackage{wrapfig}
\usepackage[english]{babel}
\usepackage{subfigure}
\usepackage{graphicx,pslatex}
\usepackage{dcolumn}
\usepackage{bm}

\begin{document}


\title{Tip-enhanced Raman imaging and spectroscopy: sensitivity, symmetry and selection rules}

\author{Catalin C. Neacsu, Samuel Berweger, and Markus B. Raschke}
\affiliation{University of Washington, Department of Chemistry, Seattle, Washington, 98195-1700}

\begin{abstract}
This article reviews the new developments in Tip-enhanced Raman scattering (TERS). 
The fundamental mechanisms underlying the Raman enhancement are discussed, including the role of the plasmonic character of the metallic tips, the nature of the optical tip-sample coupling and the resulting local-field confinement responsible for ultrahigh spatial resolution down to just several nanometers. 
Criteria for the distinction of near-field signature from far-field imaging artifacts are addressed and TERS results of molecules and nanostructures are presented. 
With enhancement factors as high as 10$^9$, single molecule spectroscopy is demonstrated. 
Spatially resolved vibrational mapping of crystalline nanostructures and determination of crystallographic orientation and domains is shown making use of the unique symmetry properties of the tip in conjunction with the intrinsic Raman selection rules. 

\end{abstract}

\maketitle

\noindent
{\bf{1. Introduction}}\\
\noindent
Optical spectroscopy provides nondestructive techniques for obtaining both structural and real time dynamic information of molecules and solids. 
Vibrational spectroscopy in particular, by directly coupling to the nuclear motion, offers insight into chemical composition, molecular bonds and their relative orientation, intermolecular coupling and zone-center phonons in crystalline solids \cite{vibspec,kuzmany_1998}. 
In that respect IR and Raman spectroscopy provide complementary information, with the feature of Raman spectroscopy of fewer constraints in terms of selection rules, readily providing access to low frequencies vibrations \cite{wilson,hendra68,hendra_a1970,yu}, and being carried out in the vis-near IR spectral range in a comparably simple experimental design. 
However, with scattering cross-sections of $\sim$ 10$^{-27}$~-~10$^{-30}$ cm$^2$, the Raman response is weak, generally requiring probing a large molecular ensemble or bulk solids \cite{aroca, demtroder}. 

It is highly desirable to combine the intrinsic chemical specificity of Raman spectroscopy with optical microscopy for the investigation of the spatial heterogeneity and composition of the analyte. 
In that regard the optical far-field Raman microscope has become an established tool for material characterization on the micrometer scale and in a confocal implementation with spatial resolution down to just several hundred nanometers\cite{tabaksblat92}. 
However, for most applications the desired spatial resolution needed often exceeds the resolution imposed by far-field diffraction \cite{abbe73,rayleigh96}.

Near-field Optical Microscopy (SNOM) provides access to higher spatial resolution \cite{lewis83, pohl84,lewis84, fillard, courjon, kawataa, novotny06, novotny} and the aperture-based SNOM using tapered glass fiber tips has been employed for nano-Raman spectroscopy \cite{jahncke95, jahncke96, webster98,serio06}. 
However, the low optical throughput of the aperture probes (10$^{-3}$ - 10$^{-5}$) severely limits the spatial resolution and the sensitivity that can be obtained, resulting in a long imaging time and parasitic Raman signal from the glass tip that could be an impediment \cite{webster98}. 
 
High sensitivity in Raman scattering, in general, can be achieved by Surface-enhanced Raman spectroscopy (SERS) providing a strongly enhanced Raman response from molecular adsorbates on rough metallic surfaces or colloidal aggregates \cite{fleischmann74, jeanmaire77, albrecht77}. 
SERS is due to the near-field enhancement of the electromagnetic field at single or coupled metal nanostructures, often resonantly excited at their surface plasmon polariton (SPP) eigenmodes \cite{moskovits78,gersten80,gersten80a,gersten80b,mccall80,kerker80, gersten81,willets07}.  
Together with a corresponding but weaker ($\sim$ 10$^1$ - 10$^2$) chemical contribution \cite{otto92} originating from surface bonding or charge transfer, the electromagnetic enhancement leads to a total increase in Raman signal by up to 14 orders of magnitude, allowing for detection down to the single molecule level \cite{kneipp97,nie97,xu99, bjerneld00,weiss01,dieringer07}. 
Despite its potential for chemically specific detection of minute amounts of analytes, it has remained challenging to develop SERS into a a routine analytic spectroscopic tool, mostly due to difficulties associated with the reproducible fabrication of SERS-active substrates \cite{michaels00,xu00,kneipp99,haynes05}. 

In general, better control over the SERS active sites and their field enhancement can be achieved by what may be viewed as resorting to an inverse geometry with respect to SERS: suspension of the metal nanostructure providing the field enhancement at a small distance above the analyte \cite{aravind82}. 
This is the basis of tip-enhanced Raman scattering (TERS) making use of a single plasmon-resonant metallic nanostructure provided in the form of a scanning probe tip of suitable material and geometry. 

Fundamentally, TERS is a variant of apertureless near-field optical microscopy (\emph{s}-SNOM) \cite{zenhausern94, inouye94,wessel85,fischer89,specht92}. 
All-optical resolution down to just several nanometers is provided by \emph{s}-SNOM, in the visible \cite{bachelot95,koglin97,sanchez99} and IR spectral regions \cite{keilmann96,knoll99,knoll00,raschke05}. 
TERS  is the extension of this technique to inelastic light scattering, with the metallic tip used as active probe which provides both the local-field enhancement and serves as efficient scatterer for the Raman emission.

\emph{s}-SNOM and special aspects of TERS have been addressed in recent reviews \cite{kawatab,novotny06,novotny,kneipp1,rasmussen05,novotny07} (and references therein), but no comprehensive discussion of the underlying physical mechanisms has yet been provided.
Here, we review our recent progress in TERS and contribution to the understanding of near-field Raman enhancement and sensitivity, the tip-sample coupling, the spatial resolution, and we underline the importance of the plasmonic character of the tip and tip fabrication. 
In addition, we show that the unique symmetry properties of the tip-scattering geometry in combination with the Raman selection rules  allows for the determination of crystallographic information on the nanoscale.
This, together with the results of other groups, shows the potential of TERS as a nano-analytical tool with diverse applications in material and surface science, and analytical chemistry for the study of biomolecular interfaces, molecular adsorbates, nanostructures and nanocomposites. \\

\noindent
{\bf{2. Tip-enhanced Raman spectroscopy (TERS)}}\\
\noindent
TERS combines the advantages of SERS with those offered by \emph{s}-SNOM: the single nanoscopic tip apex provides the local field enhancement at a desired sample location without requiring any special sample preparation  \cite{krug02,richards03}.
With the spatial resolution limited only by the tip apex size, chemical analysis on the nanometer scale is made possible.
By raster scanning the sample, spatially resolved spectral Raman maps with nanometer resolution can be obtained simultaneously with the topography in atomic force microscopy (AFM) or surface electronic properties in scanning tunneling microscopy (STM). 

The origin of the field enhancement at the tip apex is attributed to the singular behavior of the electromagnetic field (akin to the lightning-rod effect).
In addition, the spatial confinement allows for the possible excitation of localized surface plasmon polaritons (tip-plasmons) for certain tip materials \cite{kawatac}. 
With the first effect being geometrical in origin, its magnitude is mainly dependent on the curvature of the apex. 
Taking advantage of the excitation of tip-plasmons can increase the overall enhancement by several orders of magnitude as will be discussed below. 

Ultra-high sensitivity and nanometer spatial resolution imaging using TERS were obtained on various materials and molecular systems adsorbed on both flat and corrugated surfaces \cite{stoeckle00, anderson00, hayazawa00, hayazawa01, hayazawa02, anderson02, hartschuh02, hartschuh03, watanabe04, ichimura04, hayazawa04a,pettinger05,anderson05,domke06,verma06, neacsu06,steidtner07,hayazawa06,neugebauer06, zhang07}. 
Having large Raman cross-sections, several dye molecules (\emph{e.g.}, malachite green, rhodamine 6G, brilliant cresyl blue)  were used and near-field Raman enhancement factors up to $\sim$ 10$^9$ were achieved \cite{anderson02, hayazawa00,pettinger05, neacsu06,steidtner07}.
Using Ag coated AFM tips, spatial resolution below 50 nm was obtained on surface layers of Rhodamine 6G dye molecules \cite{anderson02,hayazawa00}. 
Lateral resolution as high as 14 nm and a maximum Raman enhancement factor estimated at $\sim$ 10$^4$ were obtained in spatially resolved probing vibrational modes along individual carbon nanotubes \cite{hartschuh02,hartschuh03,anderson05}. 

In studies of adenine as well as C$_{60}$ molecules, the tip-induced mechanical force was shown to lead to mechanical strain induced frequency shifts of the normal Raman modes  \cite{watanabe04,verma06}. 
Furthermore, it was observed that when interacting with individual metal atoms of the tip apex, adenine molecules form different isomers, demonstrating the potential TERS for atomic site selective sensitivity \cite{hayazawa06}.

Extension of TERS implementation for coherent spectroscopy was shown for Coherent Anti-Stokes Raman scattering (CARS) of adenine molecules included in a DNA network
\cite{ichimura04}. 
Owing to the third order nonlinearity of the CARS process, the induced polarization at the tip apex is further confined, and higher lateral resolution is in principle possible \cite{hayazawa04a}. 

Concomitant, theoretical studies on TERS report field enhancements up to three orders of magnitude in particular frequency regions \cite{jersch98,demming98,klein02,mills02,wu02,micic03}.
However, the expected resulting TERS enhancement of 12 orders of magnitude has not yet been observed experimentally.

In recent work from our group, we have refined the metallic tip fabrication, and experimentally identified and theoretically discussed the importance of the plasmonic properties of the scanning tip for achieving high Raman sensitivity \cite{neacsu05,neacsu05a,roth06, behr08}. 
This has enabled near-field Raman enhancement factors of up to 10$^9$ from malachite green molecules adsorbed on smooth Au surfaces to be obtained, allowing for the detection of TERS response with single molecule sensitivity \cite{neacsu06, neacsu07b}.

The review is organized as follows:
The experimental arrangement is presented in section 3. 
This includes the laser excitation and Raman detection, the metallic tip fabrication by electrochemical etching and the molecular systems used. 
Section 4 discusses the experimental characterization of the optical properties of the tips including their plasmonic behavior and the local-field enhancement factor as determined by second harmonic generation. 
Section 5 describes the theoretical analysis of the near-field distribution at the tip apex together with its spectral characteristics. 
The tip-sample optical coupling is discussed in section 6, where its effect on sensitivity, spatial resolution and spectral shift of the plasmon resonance are derived. 
The near-field character vs. far-field imaging artifacts in TERS and its polarization dependence are addressed in section 7.
The procedure for estimating the near-field enhancement factor will be detailed and representative values are discussed.
Tip-enhanced near-field spectra of monolayer and sub-monolayer of molecular adsorbates on smooth Au surface are given in section 8. 
The high sensitivity obtained and the dependency of the spectral features on the enhancement level is discussed. 
In section 9 we also address the important question of molecular bleaching and possible chemical contamination paths and show a number of control experiments.
Near-field tip-enhanced Raman results with single molecule sensitivity are shown in section 10. 
This is concluded from the ultra-low molecular coverage and the observed intensity and spectral temporal fluctuations.
We identify and propose in section 11 a new and promising extension of TERS for determination of both chemical and structural properties of nanocrystals.
Section 12 gives an outlook on TERS, and novel ways to circumvent current instrumental difficulties are discussed.\\

\noindent
{\bf{3. Experimental}} \\
\noindent
Various experimental schemes have been employed for TERS experiments. 
A tip axial illumination and detection geometry has been used, allowing for high NA, but requiring transparent samples or substrates \cite{hartschuh03,anderson_nl2006}. 
Similarly, but allowing to probe non-transparent samples, a high-NA parabolic mirror can be used \cite{debus_jm2002,anger_jm2003}.
In both schemes, the tip is illuminated along the axial direction, with the tip apex positioned in the laser focus.
For these geometries, polarization conditions require either a Hermite-Gaussian beam \cite{novotny98} or radial incident polarization \cite{novotny_prl2001,debus_jm2002}.
While allowing for efficient excitation and detection with the tip, independent polarization and k-vector control is limited, but desirable for symmetry selective Raman probing.

In contrast, side-on illumination and detection allows for greater flexibility in the selection of polarization and k-vector as well as the use of transparent samples.
The scanning and tip-sample distance are controlled using either AFM or STM, with STM restricted to the use of conducting samples. 

\begin{figure}
\centering
\includegraphics[width=5.5 cm]{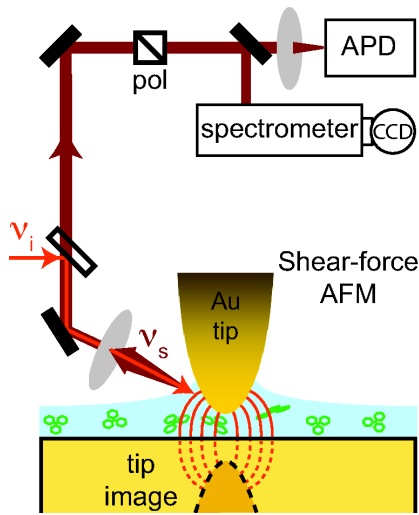} 
\caption{
Schematic of the experimental  TERS arrangement.
The incident light is focused onto the tip-sample gap. 
The tip-backscattered and -enhanced Raman response 
is spectrally filtered using a notch filter and is spectrally resolved by an imaging spectrometer with N$_2$ (l) - cooled CCD or integrally detected by means of an avalanche photodiode (APD).
Polarization directions of both incident and scattered light beams can be controlled independently. 
The blue layer indicates a thin surface water layer on the sample.}
\label{fig1}
\end{figure}
Fig.~\ref{fig1} shows the experimental layout of our side-illuminated TERS experimental arrangement.
The incident radiation ($\nu_i$) is focused onto the tip-sample gap and the tip-scattered Raman light ($\nu_s$) is detected. 
For the experiments described here, a shear-force AFM is used. 
Unlike dynamic non-contact AFM, shear-force AFM maintains a constant height of several nm above the sample.
Due to the short range tip-sample distance dependence of the field enhancement, dynamic non-contact AFM is less suitable.
The time-averaged signal is greatly reduced due to the oscillating tip action. 
Contact AFM, maintaining constant and small tip-sample distance, experiences strong forces, making it unfavorable for probing molecular or soft matter samples.
In contrast, with the spatial range of shear-forces confined to within $25$nm \cite{karrai00}, the shear-force AFM tip is controlled in close proximity to the sample without actual physical contact.

The control mechanism in shear-force AFM is based on the near surface vibrational damping of a probe tip oscillating parallel to the surface.
The nature of the shear-force damping mechanism is not yet fully understood \cite{novotny}, with a variety of mechanisms being discussed \cite{stipe01, gregor96, williamson_um1998, hoppe_um2005}.
It has been suggested \cite{karrai95,davy_apl1998} that the tip experiences viscous damping from from a thin water layer adsorbed on the surface of the sample under ambient conditions \cite{davy_apl1998,okajima_or1998,durkan_jap1996}.
This water layer, present on most hydrophilic samples, may play an important role in the surface diffusion of the analyte molecules and possibly transition of molecules to adsorb onto the tip.

As incident light source, a continuous wave Helium-Neon laser, with $\lambda_i = 632.8$nm ($1.92$eV) is commonly used \cite{kuzmany_1998}.
In our experiments, after passing through a laser-line filter, the light is focused onto the tip-sample gap by means of a long working distance microscope objective (NA = 0.35).
The tip-backscattered light is collected with the same objective and spectrally filtered using a notch filter and the signal is detected using either an avalanche photodiode or spectrally resolved using a fiber-coupled imaging spectrograph with a N$_2$(l)-cooled CCD detector. 
Even for large enhancements the signal intensities are weak and detector noise is one limiting factor. We therefore limit the spectral resolution to 25 cm$^{-1}$ for the tip-enhanced experiments. 
Far-field spectroscopic studies of molecular monolayers serving as reference to quantify the enhancement are conducted using a micro-Raman confocal setup, based on an inverted microscope (Zeiss Axiovert 135).

For our experiments we chose malachite green (MG), an organic triphenylmethane laser dye with an absorption peak around $\lambda\approx$ 635 nm. 
The absorption peak of MG is very close to the laser energy used, leading to a resonant Raman excitation via the S$_0$-S$_1$ electronic transition of the conjugated $\pi$-electron system, as discussed below \cite{bernstein79}.
To limit the rate of the molecular decomposition, the maximum fluence in the focus of the microscope objective was $5 \times 10^3~-~3 \times 10^4$ W/cm$^2$. 
However, molecular bleaching prevails under ambient conditions under resonant Raman excitation in TERS \cite{neacsu06,pettinger04}. \\

\noindent
\emph{Tip fabrication}\\
\noindent
The metallic scanning probe tips hold the central function in TERS studies providing the enhanced electromagnetic field at their apex.
Ideally, as discussed below, they present strong plasmon resonances in the spectral region of interest, leading to enhanced pump ($\nu_{\rm i}$) and scattered Raman fields ($\nu_{\rm S}$) at the apex. 
Since the first experiments, the fabrication of suitable tips has been a major challenge, and a variety of methods for their fabrication have been used: angle-cutting the metal wire \cite{guckenberger}, DC or AC voltage electrochemical and milling procedures \cite{ibe90,vasile91,nam95,iwami98}, focused ion beam milling \cite{sanchez99}, metal-coating commercially available cantilever AFM  tips \cite{ichimura_jpcc2007, ossikovski_prb2007} and attaching spherical or other plasmonic nanoparticles to a glass tip \cite{kuhn:017402}.
Electrochemical etching \cite{debus_jm2002, pettinger04a} is most commonly used due to perceived advantages \cite{picardi_epjap2007} of tips fabricated in this manner.

For our experiments, we employ a DC voltage electrochemical etching method for both Au and W tips. 
It involves the anodic oxidation of the metal wire \cite{klein97}.
In the case of Au it involves the formation of soluble AuCl$_{4}^{-}$ which subsequently diffuses away from the electrode \cite{pettinger04a}.
 \begin{figure}[t]
\centering
\includegraphics[width=8.6 cm]{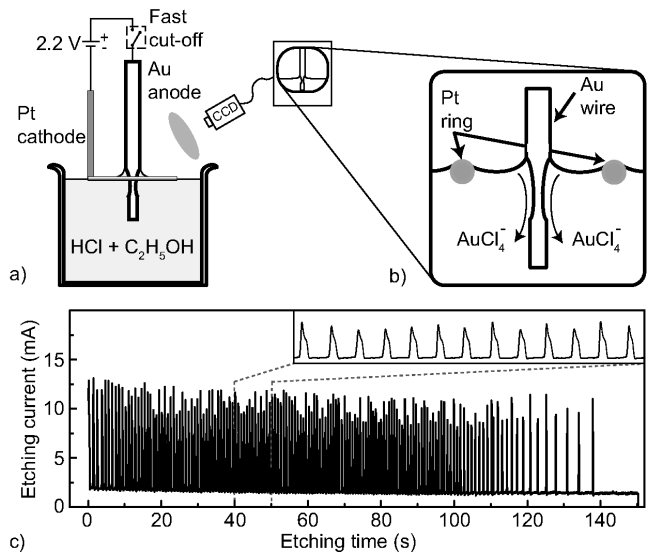} 
\caption{
Schematics of the electrochemical etching cell. 
a) The Au wire (anode) is immersed partially into the solution. 
It is surrounded by the Pt cathode-ring. 
The evolution of the process is closely followed using a video microscope. 
b) The etching takes place just underneath the meniscus formed around the Au-wire. 
The flowing of AuCl$_{4}^{-}$ is shown. 
For W-tips a similar procedure is used (see text).
c) Time evolution of the etching current for a  of 2.2 V. 
The current oscillation are periodic for about 100 seconds after which they become  less frequent, although maintaining the amplitude. 
For periodic current oscillations (inset) tips with smooth surface and consistent taper are obtained.
}
\label{fig2}
\end{figure}
The schematic of the electrochemical cell used is depicted in Fig.~\ref{fig2} panel a).
After careful cleaning with acetone the Au wire ($\phi$=0.125 mm, purity 99.99\% , temper as drawn, Advent  Research Materials Ltd.) is partially immersed ($\sim$ 2-3 mm) into the electrolyte solution. 
As electrolyte a 1:1 mixture of hydrochloric acid ({HCl}, aq. 37$\%$) and ethanol is used. 
As the cathode, a platinum wire ($\phi$ = 0.3 mm) circular ring electrode with a diameter of $\sim$1 cm is used. 
It is held at the surface of the electrolyte  with the Au wire positioned at the center of the Pt ring.
For the etching a potential of +2.2V is applied to the Au anode with respect to the Pt cathode.
This voltage was determined by us as well as others \cite{wang_apl2007} to produce the best tips.
This value is well above the Au oxidation potential due in part to the activation energy along the reaction pathway \cite{ibe90}.

When placed in the electrolyte solution, the surface tension causes a concave meniscus to form around the wire, as shown schematically in Fig.~\ref{fig2} panel b). 
The overall shape and aspect ratio of the tip after etching are primarily determined by the shape of the meniscus \cite{pettinger04a}.
During etching a downward flow of AuCl$_4^-$ along the wire can be observed.
The resulting ion concentration gradient partially inhibits etching of the lower portion of the wire, resulting in a necking of the wire near the meniscus \cite{melmed90}.
This proceeds until the lower section of the wire falls off. 
The remaining upper part of the wire is then used as the TERS/AFM tip. 

Since the tip remains in the solution under the meniscus after the detachment of the lower part, the circuit has to switch off as rapidly as possible, for further etching would result in blunt tips. 
A comparator breaks the etching voltage when the current value becomes smaller than an adjustable reference value determined from the current change associated with the drop of the lower tip.

Monitoring the etching current reveals periodic oscillations of the current, as is shown in Fig~\ref{fig2} c). 
For the potential of $2.2$V, the etching will generally reach completion in approximately 150 seconds with an average baseline current of $\sim$2mA.
After a short time of initial fluctuations, the period equilibrates and remains constant for the first $\sim$100 seconds of the etching process. 
Towards the end of the etching process, the oscillation period continuously decreases until completion of the process.

These current oscillations have been attributed to the depletion of Cl$^{-}$ near the surface of the electrode \cite {pettinger04a}.
Initially the Au will react rapidly to form AuCl$_4^-$, depleting the Cl$^{-}$ near the electrode--electrolyte interface,  resulting in a period of high current.
With decreasing local Cl$^-$ concentration, the Au will more readily form an electrode--passivating layer of Au(OH)$_3$ \cite{frankenthal_jes1976}, leading to extended periods of decreased current.
Upon restoring the Cl$^-$ concentration, the passivating layer dissolves and another current spike occurs \cite{mao_jec1995}.
Associated with the complex nature of the details of this oscillating electrochemical process, it is sensitively dependent on the applied potential. 
An empirically established etching voltage leads to the most periodic oscillations and results in the highest quality tips.

It is desirable to have a criterion to select suitable TERS tips other than scanning electron microscopy (SEM), which is known to deposit Raman - visible carbon contamination onto the tips due to electron-beam induced decomposition of trace organics in the residual gas \cite{hillier_jap1947, guise04}.
Tips etched under a constant oscillation period exhibit a smooth surface and consistent taper. 
In contrast, tips etched under conditions resulting in an inconsistent periodicity of the etching current frequently present deformities and large surface irregularities. 
We could verify in our experiments, by comparison of TERS activity with SEM of tip shape as well as the study of SPP of the tip apex, a link between homogeneous and smooth taper with TERS performance.
Our observations here are in good agreement with previous work by Wang et. al \cite{wang_apl2007}.

A similar etching procedure is used for the tungsten tips. A W wire ($\phi = 200  \mu$m) is partially immersed ($\sim$ 2-3 mm) in aqueous 2 M KOH. 
A DC voltage of 3 V is applied between the wire and a stainless steel ring cathode.

Using these procedures tips with apex radii as small as $10$nm are obtained.
After etching, the tips are cleaned in distilled water and stored in isopropanol prior to usage, to avoid possible contamination in an otherwise uncontrolled atmosphere. \\

\noindent
{\bf{4. Metallic tip optical characterization}}\\
\noindent
Efficient local-field enhancement and TERS activity is in general associated with the excitation of local modes of surface plasmon polaritons (SPP) at the metallic tip \cite{denk91,novotny97,krug02,hartschuh03, porto03}. 
However, determining the details of this correlation has remained an open problem.
In the following we present the investigation of the spectral characteristics of the elastic light scattering from individual sharp metal tips and discuss the results in the  context of the local plasmonic resonant behavior. \\

\noindent
\emph{Surface plasmon polaritons of Au tips}\\
\noindent
Dark-field scattering spectroscopy with white light illumination would lead to a largely unspecific response with the scattering dominated by the tip shaft \cite{bohren}.
\begin{figure}[t]
\centering
\includegraphics[width=7 cm]{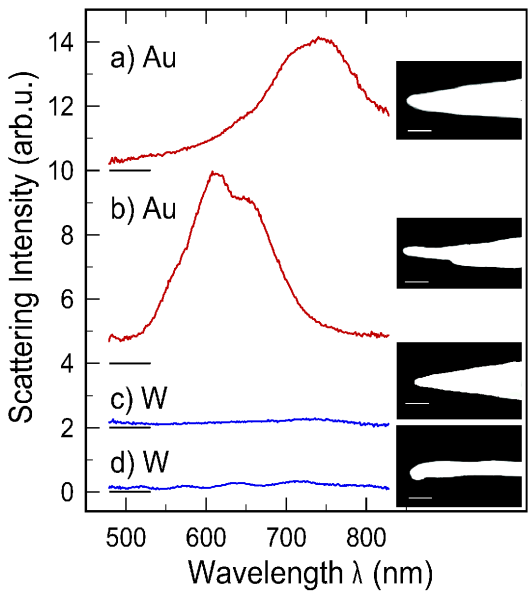} 
\caption{
Scattering spectra for different Au and W tips.
A plasmon-resonant behavior is observed for Au tips  (a, b).
Weaker and in general spectrally flat signal is observed in the case of W tips  (c, d).
The spectral data are juxtaposed with the electron microscope image from the corresponding tip.
The scale bar corresponds to 100 nm \cite{neacsu05a}.
}
\label{fig3}
\end{figure}
The SPP characteristics of the apex itself would become difficult to distinguish. 
Therefore, for the plasmonic light tip-scattering experiments we spatially limit the optical excitation to the near-apex region by use of evanescent wave excitation.
For that purpose, the tip frustrates the evanescent field formed by total internal reflection on a prism base  \cite{kretschmann72,leurgans47} and the tip-scattered light is detected and spectrally analyzed. 
This confines the excitation to just several 100 nm from the tip apex.
The complete description of the setup and the results is given elsewhere \cite{neacsu05a}.

Fig.~\ref{fig3} shows representative scattering spectra for different Au (a, b) and W (c, d) tips. 
Both the excitation and detected light fields are unpolarized. 
All spectra are acquired for the tips within few nanometers above the prism surface, as controlled by shear-force AFM. 
The intensity scale is the same for all four cases, and the spectra are offset for clarity.
Electron micrographs for the tip structures investigated are shown as insets.

The pronounced wavelength dependence of the scattering of Au tips is characteristic of a plasmon resonant behavior. 
Both scattering intensity and spectral position of the resonance are found to be sensitive with respect to structural details of the tips.
In general, for regular tip shapes the resonance is characterized by one (Fig.~\ref{fig3} a) distinct spectral feature. 
Inhomogeneities in the geometric shape are reflected in spectral broadening and/or occurrence of multiple spectral features (Fig.~\ref{fig3} b).
In addition, the spectral position and shape of the plasmon resonance depend sensitively on the aspect ratio of the tip. 

For comparison, spectral light scattering by tungsten tips of similar dimensions is shown in Fig.~\ref{fig3} c and d.
Overall weaker emission intensities are observed compared to Au tips. 
For W, a metal with strong polarization damping due to absorptive loss in the visible and near-IR region, no SPP resonance is expected. 
Here, a spectrally flat optical response is observed with weak overall scattering intensities (Fig.~\ref{fig3} c).
The spectral behavior is found to show little variation with tip radius and tip cone angle, except for the case of very slender tips where a modulation is observed, as shown in Fig.~\ref{fig3} d). 

\begin{figure}[t]
\centering
\includegraphics[width=8.6cm]{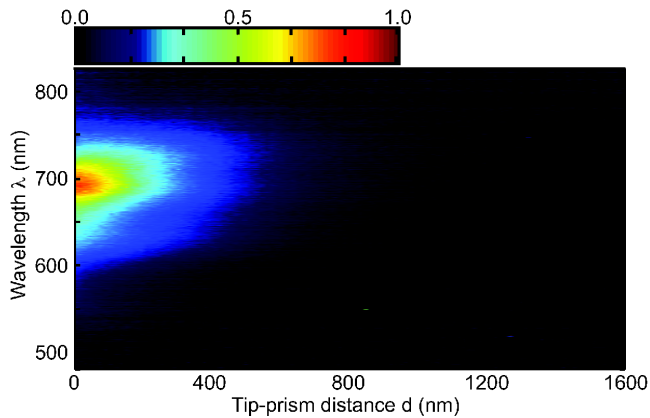} 
\caption{
Spectrally resolved elastic scattering response of a Au tip approaching the evanescent white light generated at a prism base. 
A surface plasmon resonance centered around $\lambda$ = 690 nm is observed. 
The absence of a spectral shift associated with the tip approaching the prism surface demonstrates the tip-apex region as the origin of the resonance peak.
}
\label{fig4}
\end{figure}

A spectrally resolved tip-prism distance dependence measurement of the scattering intensity for a Au tip with apex radius of r $\sim$ 15 nm is shown in Fig.~\ref{fig4} with the plasmonic resonance peak centered around $\lambda$ = 692 nm. 
The scattered signal is normalized with respect to the evanescent source spectrum that varies as a function of distance \cite{fornel}. 
An increase in scattered intensity with decreasing the tip-prism distance is seen as the tip frustrates rising evanescent field intensities.
The absence of a change in the spectral scattering characteristics demonstrates that the signal intensity observed at the strong resonance is dominated by the tip-apex region, with the shaft contributing to a spectrally unspecific background.

From polarization dependent studies of the tip scattering process we have found that, in general, more intense scattering is observed for emission polarized parallel with respect to the tip axis ($p$-polarized emission), corresponding to the excitation of longitudinal plasmonic modes.
The intensity ratio of $p$ to $s$ (emission orthogonal to the tip axis) is typically found to range between 2 to 5 with a maximum value of $\sim 10$. 
In the context of the TERS experiments it must be noted that the $s$-polarized field is not expected to get enhanced near a surface. 
In this case, the optical polarization of the tip and the corresponding image polarization induced in the sample are oriented antiparallel, as discussed below.  \\

\noindent
\emph{Local-field enhancement from bare tips}\\
\noindent
The highly resonant characteristics observed for Au tips suggest strong local-field enhancements in the vicinity of the apex. 
The quantification of the near-field enhancement factor is a difficult task in general. 
Without an absolute reference, the enhancement factor cannot be quantified from these linear optical experiments. 
We therefore make use of the symmetry selectivity of the second order nonlinear optical response in the form of second harmonic generation (SHG) from the apex region of the tips. 

SHG is forbidden in the dipole approximation for media with inversion symmetry \cite{heinz85}. 
In scattering geometry, for sagittal illumination and tip-parallel polarization, the SHG response from the tip is then dominated by the apex region, where the macroscopic translational invariance is broken in the axial direction. 
With the symmetry being radially conserved, little signal is expected from the conical near-apex shaft area \cite{note2}.
For the experiments, linearly polarized incident light from a mode-locked Ti:sapphire oscillator (pulse duration $<$ 15 fs, $\lambda$ = 805 nm) is directed onto the sharp end of the free standing tip and the scattered SHG signal is spectrally selected and detected. 
 
The contribution of the local field-enhancement of SHG from the metal tips is derived comparing the signal strength obtained with that of a planar surface of the same material. 
With the SH-enhancement expected to be dominated by the tip apex, a SH-enhancement of $\sim 5\times10^3 -4\times10^4$ was observed for Au tips with $r \simeq 20\,{\rm nm}$.
With the SH-power $\propto E^4$ \cite{boyd} this corresponds to an amplification of 8 -- 25 for the average electric field near the apex, in agreement with estimates based on other SHG experiments  \cite{ropers08}.
For W tips significantly lower values for the SH enhancement are found corresponding to local field factors between 3 and 6. 
These results are also in good agreement with theoretical models, despite microscopic variations in the details of the tip geometry, as discussed below.

The excitation of the localized surface plasmon polariton (SPP) in the axial direction is responsible for the field enhancement observed experimentally for Au tips.
A systematic investigation of the influence of these geometric parameters in terms of cone angle and tip radius would be highly desirable; however, the limitations due to the electrochemical preparation procedure render this difficult.\\

\noindent
{\bf{5. Calculation of the near-field distribution at the tip-apex}}\\
\noindent
The near-field distribution and enhancement has been derived for a variety of tip model geometries and tip and sample material combinations using different theoretical methods \cite{bouhelier03a, denk91, porto03, festy04, schneider05, neacsu05, neacsu06, goncharenko06, behr08} (and references therein).
The accurate theoretical treatment of the problem involves the solutions of MaxwellÕs equations.
This can be performed numerically for a chosen model tip-geometry \cite{martin01, downes06, novotny97,roth06}. 
Although this may closely reproduce the experimental observations, the approach is computationally very demanding. 
It has remained difficult to extract the underlying relevant microscopic parameters responsible for the optical response observed, given that the effects of tip geometry, tip material, tip-sample distance, and optical field are coupled. 

Taking advantage of the small dimensions of the tip apex compared to the optical wavelength ($kr \ll 1$, with $k$ the wave vector and $r$ the tip-apex radius), the problem can be treated in the quasistatic approximation which allows solving the Laplace equation analytically for certain geometries \cite{denk91,porto03,festy04, schneider05, goncharenko06, behr08} to derive the local field distribution \cite{stratton,demming98}.
With the size of the apex region with $r \sim$ 10 nm this implies that the electric field has the same amplitude and phase across the structure at any time and thus retardation effects can be neglected \cite{metiu84, bohren}.
Despite constraints in terms of tip geometries which can be treated in that approach, in contrast to purely numerical techniques this method provides direct insight into how the solutions scale with several  experimentally relevant structural and material parameters. 

\begin{figure}[t]
\centering
\includegraphics[width=8.3cm]{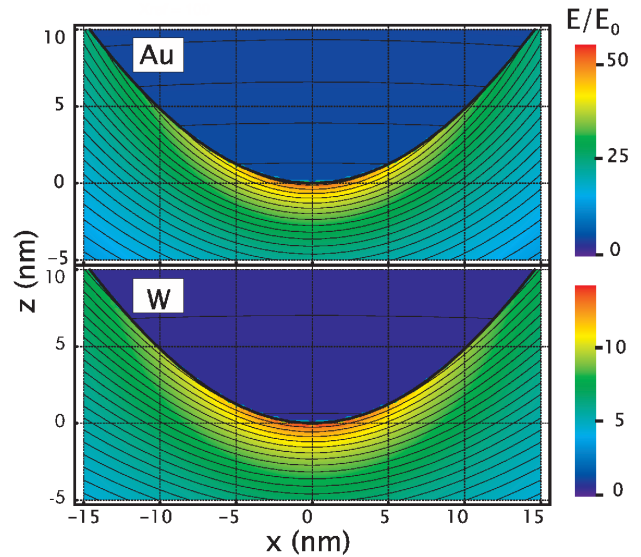} 
\caption{
Dependance of field enhancement ($E/E_0$) on tip material for free standing tips with apex radius $r$ = 10 nm, and wavelength of $\lambda$ = 630 nm. 
The solid lines represent contours of constant potential \cite{behr08}.}
\label{fig5}
\end{figure}

Here, we represent the tip as a hyperboloid and the influence of  different dielectric and structural parameters on the near-field enhancement and distribution is systematically derived for both bare tips and tip-sample systems \cite{behr08}. 
The optical wavelength dependance is explicitly taken into account considering the frequency dependence of the dielectric functions of tip and sample media \cite{palik}. 

Fig. \ref{fig5} shows characteristic local field distributions and the corresponding enhancement near the apex region of free standing tips of gold (a) and tungsten (b). 
The equipotential surfaces are indicated by solid lines.
In both cases the radius of the apex and the cone semiangle are fixed to $r$ = 10 nm and  to $\theta$ = 20$^\circ$, respectively. 
Setting the incident light at a wavelength of $\lambda$ = 630 nm and polarized along the tip axis, this closely resembles the conditions in the TERS experiments. 

At optical frequencies, even for metals as tip material, the tip surface does not represent an equipotential surface, in contrast to the pure electrostatic case. 
The finite response time of the charge carriers with respect to the optical frequency results in the decay of the field inside the tip on the length scale given by the skin depth. 
For gold as a representative material with high conductivity this results in the strongest field enhancement of $E/E_0$ $\approx$ 50 at the apex \cite{note1}. 
In contrast, tungsten as a common scanning probe tip material, is a poor conductor in the optical frequency range leading to a comparably moderate enhancement of $\sim$ 12. 
The degree of field enhancement $E/E_0$ at the tip apex depends sensitively on apex radius and cone semiangle, due to their influence on the plasmon resonance as discussed briefly below and in \cite{behr08}.
In general, values range between 10 and 100 for typical gold tips with 10 -- 20 nm radius and realistic semiangles.

Despite the necessary approximations inherent to the quasistatic approach, the theoretical results presented here prove to be sufficiently accurate for most practical purposes. 
This is drawn from comparison with the experimental results shown above. 
From the tip-scattered SHG experiments the local field enhancement of 8 -- 25 for Au and 3 -- 6 for W was estimated for  $r$ = 20 nm apex radii.
Considering that the experimental enhancement factors presented above are obtained as an spatial average over the apex region, these values fall well within the range of the theoretically predicted enhancements given in Fig.~\ref{fig5}.\\

\noindent
{\bf{6. Tip-sample optical coupling}}\\
\noindent
The local field enhancement as well as the lateral confinement can change significantly for the tip in close proximity to a surface plane. 
This behavior is of crucial importance for the optical contrast in scattering near-field microscopy. 
The optical tip-sample coupling is the result of the forcing of the boundary conditions at the surface plane on the field emerging from the apex. 
With the incident electric field inducing an optical dipole excitation in the tip, the presence of the sample can be accounted for by considering a virtual image dipole located inside the sample, with the resulting field distribution being a superposition of the fields of the two dipoles \cite{jackson}.
This gives rise to a mutual and constructive tip-sample optical polarization when the electric field is oriented parallel with respect to the tip axis ($p$-polarized). 
For an $s$-polarized incident field, the tip - dipole is induced parallel to the sample surface, and the correspondent image - dipole aligned antiparallel. 
This leads to a partial cancellation and hence reduced field intensity and scattering	 \cite{raschke05}.

Fig.~\ref{fig6} (top panel) displays the evolution of the field in the tip-sample gap calculated along the axial direction for different distances $d$ for a Au tip approaching a flat Au sample.
As can be seen, the tip-sample approach is accompanied by a significant increase in field enhancement in the tip-sample gap. 
The tip-sample interaction is correlated with apex radius and becomes significant at distances below about twice the tip radius (here, $d$ = 20 nm) when the near-field interaction becomes effective,  with a particularly fast rise of the field at the sample surface. 
It reaches values of up to several hundred for $d$ = 2 nm showing an increase of more than one order of magnitude when compared with the free standing tip. 
In addition, the field enhancement with decreasing tip-sample distance is accompanied by a strong lateral confinement of the field underneath the apex \cite{behr08}. 
The equipotential surface is forced to align close to parallel with the substrate plane which gives rise to an enhanced lateral concentration of the field. 

The enhancement is strongly dependent on the sample material, being most pronounced for metallic substrates. 
Simultaneously, the lateral confinement also varies with the sample, {\emph{i.e.}, decreases with decreasing the material optical polarizability of the material. 
This is important as it leads to an increased spatial resolution in scanning probe near-field microscopy for small tip-sample distances, as discussed in \cite{behr08}.

\begin{figure}[t]
\centering
\includegraphics[width=8.3 cm]{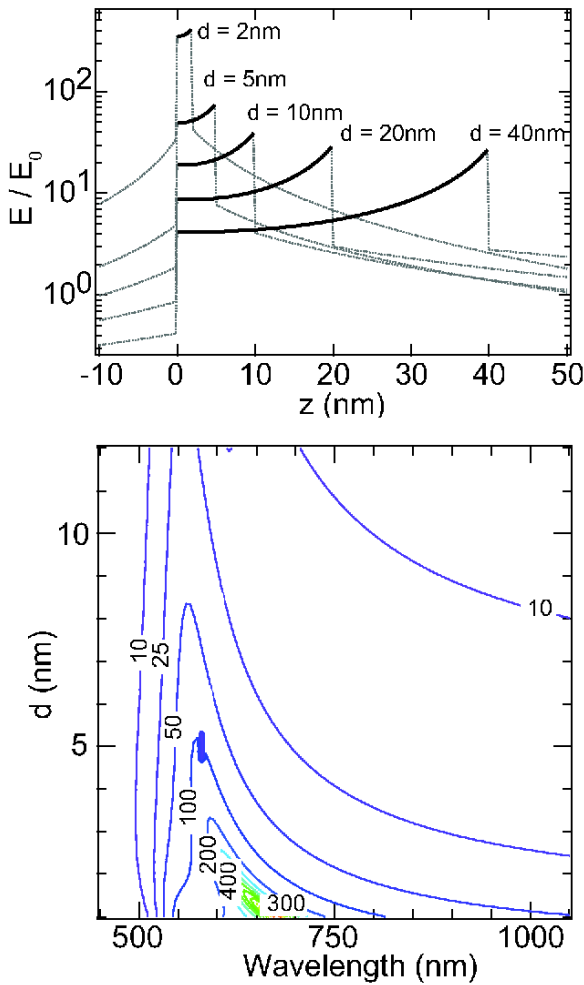} 
\caption{
Top:Variation of field enhancement $E/E_0$ along the axial direction across the tip-sample gap region for different distances $d$ for a Au tip ($r$ = 10 nm) and Au sample at an excitation wavelength of $\lambda$ = 630 nm. 
The tip is at variable $z$ = $d$ positions and extends to the right. 
The sample surface is located at $z$ = 0 nm and its bulk occupies the range of negative z-values.
Bottom: Spectral dependance of field-enhancement with tip-sample distance d for a Au tip ($r$ = 10 nm) approaching a Au surface. 
The pronounced red shift of the plasmon response is associated with the strong near-field tip-sample coupling for $d\leq r$. 
The lines represent contours of equal optical field enhancement \cite{behr08}.}
\label{fig6}
\end{figure}

One of the virtues of the quasistatic model is the direct access to the spectral variation of the field distribution for different tip-sample geometries.
The spectral tip-scattered response can become a complex superposition of the tip and the sample optical properties, the understanding of which is important in nano-spectroscopy.

 For a Au tip ($r$ = 10 nm) approaching a Au surface Fig.~\ref{fig6} (bottom panel) shows the calculated spectral dependance of the field enhancement near the surface. 
 As expected, a structural plasmon resonant behavior is observed. 
 Associated with the increase in field enhancement for shorter distances a spectral shift in the plasmon response to longer wavelengths is observed. 
 This red shift is especially pronounced for distances $d\leq r$, {\emph{i.e.}, correlated with the onset of the sharp rise in field enhancement as discussed above and a manifestation of the regime of strong coupling. 
Spectral peak widths of order 0.2 to 0.3 eV correspond to what is expected from the electronic dephasing times for SPP in Au of $\sim 10-20$ fs \cite{raether}. 
Using tungsten as tip material, no plasmon behavior is obtained, except for the case when it is combined with a sample that can sustain an SPP itself \cite{behr08}.

The determination of the field enhancement for a tip-sample coupled system has been been experimentally achieved by TERS from surface monolayers of molecular adsorbates \cite{hartschuh03, neacsu06, roth06, pettinger05, zhang07}. 
From tip-sample distance dependent Raman measurements in comparison with corresponding far-field experiments, using malachite green (MG) dye molecules or single-walled carbon nanotubes (SWCN) near-field enhancements of 60 - 150 at the sample surface were measured, as detailed below. 
These experimental values are in good agreement with the theoretical ones ranging from $\sim$ 50 to $\sim$ 300 as shown in Fig.~\ref{fig6} for small tip - sample distances.

The spectral characteristics of the plasmon response found in these calculations and shown in Fig.~\ref{fig6} display a red-shift for the case of a Au tip approaching a Au surface. 
It is the result of the superposition of the dielectric functions of the tip and the sample material mediated by the tip-sample optical coupling.
This is a general phenomenon and it is found in calculations of spheres and other plasmonic nanostructures in close proximity to a metal surfaces \cite{downes06, porto03, geshev04, aravind81}, and has been observed experimentally in TERS and light emission in inelastic tunneling \cite{pettinger07, berndt91,aizpurua00}. 

The field distributions and enhancement and their spectral dependence calculated within the quasistatic approximation for a hyperbolical tip are found to agree with other detailed theoretical observations. 
Using a fully 3D finite-difference time-domain (FDTD) method we have calculated the field distribution \cite{roth06} with similar results obtained in \cite{festy04,demming05, downes06}.
The comparison with both the exact theoretical treatments and experimental results validates the approach of treating the probe tip in the quasistatic approach to a good approximation.
Despite the simplicity of the model, the essential optical properties and the physical trends characteristic for the optical response of the the tip-sample system are accurately predicted.\\

\noindent
{\bf{7. Near-field character and far-field artifacts in TERS}}\\
 \noindent
TERS manifests itself in an enhancement of the Raman response, with the increase confined to the region underneath the tip-apex. 
However, with the illumination extended on a larger surface region determined by the far-field focus, the discrimination of the variation of far-field response due to the presence of the tip inside the focus is difficult in general  \cite{hartschuh03}. 

Without any lateral scanning or systematic vertical tip-sample distance variations this in general does not allow for the unambiguous assignment of the observed optical effect to a near-field process. 
The apparent Raman signal rise may be due to far-field effects occurring when the tip is scanned inside the tight laser focus that can influence both signal generation and detection. 
With the tip penetrating into the focus region it would allow to scatter additional, otherwise forward-scattered (non-enhanced) far-field Raman light back into the detector. 
Furthermore, the interference of tip-scattered, surface reflected and incoming pump light results in locally enhanced pump intensities. 
With both processes affecting a surface region not confined by the apex area, they can dominate over the near-field effects.

\begin{figure}[t]
\centering
\includegraphics[width=8.3cm]{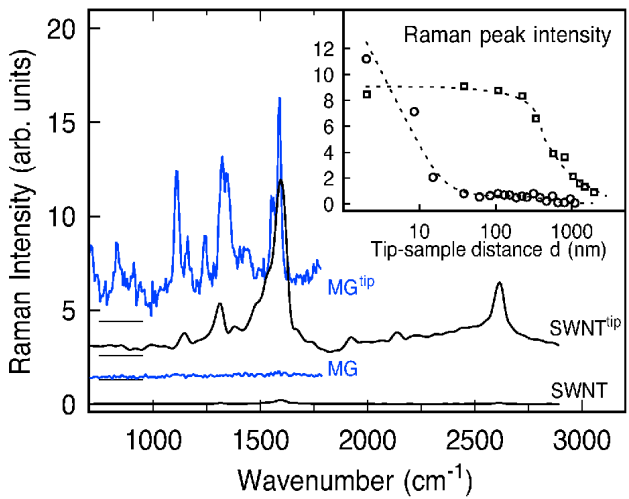} 
\caption{
Near-field signature vs. far-field artifact: 
Raman spectra of single-walled carbon nanotubes and MG molecules with tip retracted (SWNT, MG) and tip engaged (SWNT$^{tip}$, MG$^{tip}$). 
Inset: tip-sample distance dependence of the Raman signal obtained under similar conditions but displaying very different behaviors: 
the $\sim$20 nm length scale increase is characteristic for the near-field signal origin (circles); the few hundred nanometer decay length (squares) shows a far-field artifact - leading to similar signal increase as the near-field response. 
Dashed lines added as guide for the eye.}
\label{fig7}
\end{figure}

In Fig.~\ref{fig7} tip-scattered Raman results are shown for single-wall carbon nanotubes (SWCN) and and monolayers (ML) of malachite green (MG) molecules with the tip in force feedback at $d=0$ nm \cite{note}, versus tip retracted by several 100 nm.  
When the tip is within several nanometers above the sample surface, a strong increase in Raman intensity is observed for both adsorbates (spectra denoted SWNT$^{tip}$ and MG$^{tip}$).
Note that the difference in noise level from the SWNT to the MG spectra is due to different spectral resolution settings of the spectrometer.
Despite being frequently applied to assign the observed signal to TERS \cite{pettinger04} simply comparing surface vs. tip scattered Raman intensities near- and far-field processes are \emph{a priori} indistinguishable. 
The inset of Fig.~\ref{fig7} shows the Raman peak intensity as a function of the tip-sample distance obtained in two similar experiments for monolayers of MG molecules adsorbed on flat Au surface. 
The overall increase in signal is comparable in both cases, and an estimate of the Raman enhancement factor gives G $>$ 10$^6$ (\emph{vide infra}). 
However, with the distance variation occurring on a length scale correlated with pump wavelength or focus dimensions, in one case the enhancement can solely be attributed to far-field effect (squares).
A true near-field effect manifests itself in a correlation of the spatial signal variation with the tip radius ($\sim$ 20 nm). 
Here, with high quality tip (sharp apex, smooth tip shaft), the near-field contribution can dominate the overall signal (circles). 
Therefore for the tip-scattered Raman signal only the demonstration of a clear correlation of the lateral or vertical tip-molecule distance dependence with tip radius allows for an unambiguous near-field assignment of the optical response \cite{hartschuh03}, as is true for all near-field microscopies including \emph{s}-SNOM and the special case of TERS \cite{novotny, kawatab, kawatac, courjon}.\\

\noindent
\emph{Experimental quantification of the near-field Raman enhancement}\\
 \noindent
In contrast to SERS where the quantification of the enhancement is a difficult task in general, for the tip-scattering experiment, the Raman enhancement factor can be derived from comparison of tip-enhanced versus far-field response of the same surface monolayer. 
 \begin{figure}[t]
\centering
\includegraphics[width=8.3cm]{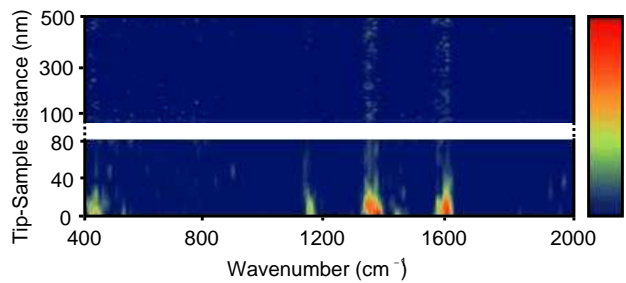} 
\caption{
Tip-sample distance dependence of spectrally resolved Raman signal during approach of $\sim$ 1 ML of MG on gold.
Each spectrum is acquired for 1 s and the approach is realized with 2 nm increments.
Near-field tip-enhanced signal is observed with the tip within $\sim$ 20 nm above the sample, displaying typical Raman modes for MG molecules \cite{neacsu06}.}
\label{fig8}
\end{figure}
Fig.~\ref{fig8} shows the spectrally resolved tip-scattered Raman signal during approach of $\sim 1$ ML of MG on gold (2 nm/step, 1 s/spectrum acquisition time).  
The pump light is polarized along the tip axis (p$^{in}$) and the Raman signal is detected unpolarized.
Although a faint Raman signature of the molecules is observed with the tip at $d >$ 100 nm, a clear molecular fingerprint is obtained only when the tip is within $\sim$ 20 nm from the sample. 
The prominent bands around 1615 cm$^{-1}$ and 1365 cm$^{-1}$ are assigned to combinations of the C=C stretching vibrations of the phenyl ring and the mode at 1170 cm$^{-1}$ is due to a methyl group rocking mode or an in-plane C-H bending mode of the phenyl ring \cite{lueck93}. 

The enhancement is confined to a tip-sample spacing of just several nanometers and correlated with the apex radius of the tip, as expected for the near-field signature. 
The increase in Raman response is accompanied by a weak rise in a spectrally broad fluorescence background that has been subtracted. 
With the molecular fluorescence being quenched due to the electronic coupling to the metal substrate, this emission could largely be attributed to the enhancement of the intrinsic tip luminescence  in conjunction with excitation of plasmonic modes in the tip-sample cavity \cite{beversluis03, pettinger07}, \emph{i.e.}, its origin is independent of the molecular adsorbates.
 
With the near-field character of the Raman response verified, the experimental field-enhancement factor can be derived from comparison of the tip-enhanced versus far-field response from the same surface monolayer. 
For the experiments presented here, the integrated Raman signal over the 1150~-~1650 cm$^{-1}$ spectral region is used after background subtraction.

As shown in Section 5, the electromagnetic near-field enhancement originates from a sample surface area approximated by the area of the tip apex \cite{hartschuh03}.
For the evaluation of the enhancement factor, the different areas probed in the near-field (TERS) and far-field (FF) cases are then taken into account.
For the TERS setup, the illumination focus has a diameter $d \simeq$ $(\lambda$ / N.A.)$\times$1.5 = 1.7 $\mu$m (the 1.5 factor accounts for the deviation of the laser beam from a perfect gaussian profile).
Considering the $\sim$70$^{\circ}$ angle of incidence of the pump light with respect to the surface-normal in our setup, the actual surface region illuminated is elongated elliptically and  larger, with a total area of $\sim$ 27 $\mu$m$^2$. 
In the case of the confocal--Raman setup (FF) used to record the far-field spectrum of MG, the same laser illuminates an area of 870 nm$^2$. 
Using the inter-atomic distances the molecule is estimated to occupy a surface area of $\sim$ 0.87 nm$^2$ (actual 3D spatial filling: 1.18 nm $\times$ 1.39 nm $\times$ 0.98 nm).  
1 ML surface coverage then corresponds to  approximately 10$^6$ molecules are in the focus of the far-field setup and only $< $ 200 are responsible for the tip-enhanced signal for a tip with 10 nm apex radius. 

In addition to the areas, the detection efficiencies for the two different experimental arrangements are considered. 
The far-field response is emitted only in half space ($\Omega=2\pi$) assuming an isotropic dipolar intensity pattern, and with N.A. = 0.9 a total of $\sim$ 27\% of the Raman signal is detected, in contrast to the TERS experiments with $\sim$ 2\% for N.A. = 0.35 and assuming the emission pattern following a cos$^2(\theta+\frac{\pi}{2}, )$ dependence ($\theta \in [0,\pi]$) . 

Taking all these factors into consideration, the Raman enhancement factor can be estimated, and is found to range from 10$^6$ to 10$^9$ with variation mostly depending on the tip used.
This enhancement is expected to be purely electromagnetic in origin. 
Here the molecule experiences the tip-enhanced local-field E$_{loc}(\nu_i)$ = L($\nu_i$) E($\nu_i$), with L($\nu_i$) the enhancement factor of the incident field E($\nu_i$).
The concomitant enhancement of the polarization P($\nu_s$) at the Raman-shifted frequency $\nu_s$ is L$^\prime$($\nu_s$). 
Therefore, the total field enhancement is given by L($\nu_i$)$\cdot$L$^\prime$($\nu_s$) \cite{boyd, neacsu05a}. 
With the Raman intensity I $\propto$ $|$L($\nu_i$)$\cdot$L$^\prime$($\nu_s$)$\cdot$E($\nu_i$)$|^2$  , the total Raman enhancement factor $G$ is given by  $G$ = $|$L($\nu_i$)$\cdot$L$^\prime$($\nu_s$)$|^2$ \cite{moskovits85}. 
Although different in general, the field enhancement factors L$(\nu_i)$ and L$^\prime$($\nu_s$) for pump and Raman polarizations, respectively, can be assumed to be similar in this case \cite{haes05,mcfarland05}. 
This is motivated by both the spectrally broad plasmonic resonance of the tip and its red-shift upon approaching the sample surface \cite{aravind83, porto03, behr08}.

\begin{figure}[t]
\centering
\includegraphics[width=7 cm]{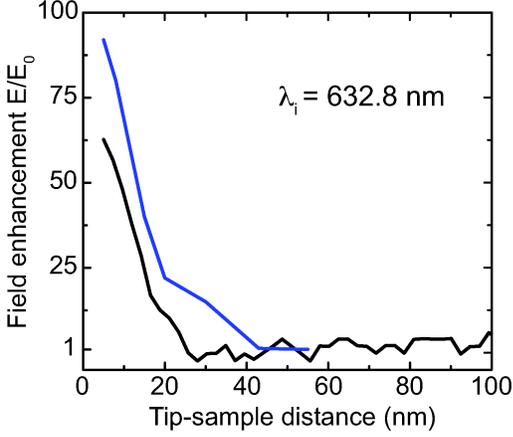} 
\caption{
Field enhancement ($E/E_0$)  for two different Au tips approaching $\sim 1$ ML of MG on gold.
The Raman response is integrated over the 1150~-~1650 cm$^{-1}$ spectral region after background subtraction and the enhancement factors are calculated according to I$_{TERS}$ $\propto$ E$^4$.}
\label{fig9}
\end{figure}
Fig.~\ref{fig9} shows the effective field enhancement factors (L($\nu_i$)$\cdot$L$^\prime$($\nu_s$) = E/E$_0$) for the integrated Raman signal from $\sim 1$ ML of MG on gold as a function of tip-sample distance for two different Au tips (both with r $<$ 15 nm). 
Maximum enhancements of $\geq$ 90 and $\geq$ 60 (black and blue curves, respectively) are obtained considering variation of the tip-scattered Raman response with the forth power of the electrical field, as indicated above.
The molecules adsorbed on the planar Au surface already experience a field enhancement given by the Fresnel factor \cite{fowles}. 
We will therefore also derive the total enhancement with respect to the free molecule response. 

The study of the polarization dependence of the Raman response offers additional insight into the electromagnetic enhancement of TERS.
Fig.~\ref{fig10} shows near-field Raman spectra from $\sim 1$ ML of MG on gold for the different polarization combinations of both pump and Raman light. 
Incident laser power and acquisition times are identical for all spectra. 
The polarization directions are defined as parallel ($p$) and perpendicular ($s$) with respect to the plane of incidence formed by the incoming wave vector k($\omega_i$) and the tip axis.
No background has been subtracted and the data are normalized with respect to the intensity of the 1615 cm$^{-1}$ mode measured in p$^{in}$/p$^{out}$ configuration (upper left panel). 

With the incident field polarized perpendicular on the tip axis (s$^{in}$), almost no Raman signal is observed, irrespective of the polarization of the scattered light.
In contrast, with the pump polarized along the tip axis (p$^{in}$), clear Raman fingerprints of MG molecules are observed with the Raman response being predominantly polarized parallel to the tip (p$^{in}$/p$^{out}$) as expected for near-field TERS from isotropically distributed molecules with diagonal Raman tensor components as the case of MG.
For both s$^{in}$/p$^{out}$ p$^{in}$/s$^{out}$ configuration, weak overall signal is observed due to the absence of the tip-sample optical coupling. 
For s$^{in}$/s$^{out}$ a larger background is observed, albeit with no Raman enhancement, as expected.

\begin{figure*}[t]
\centering
\includegraphics[width=12 cm]{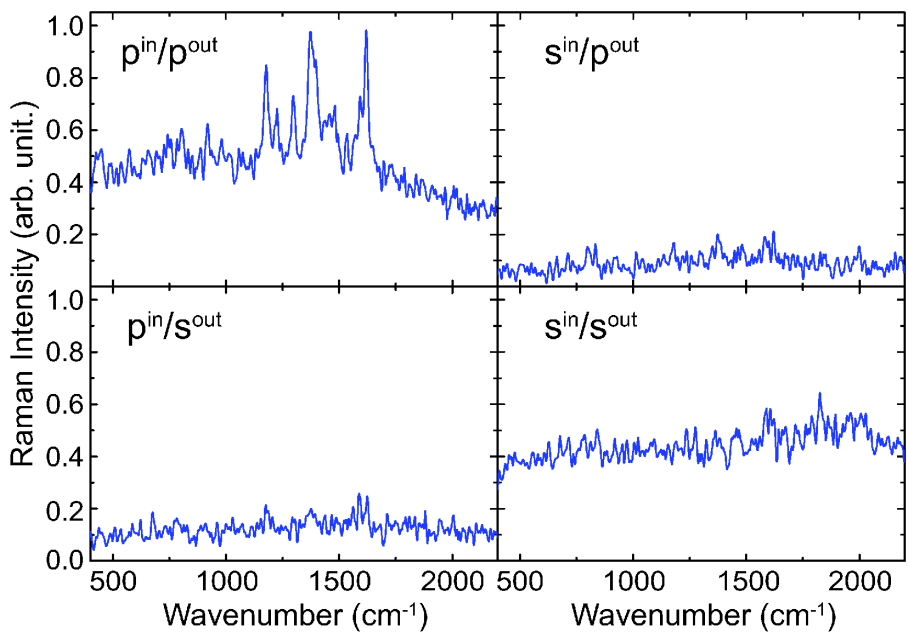} 
\caption{
Polarization dependence of the near-field tip-enhanced Raman response originating from $\sim 1$ ML of MG on gold.
All spectra are acquired for the same time and normalized to the maximum intensity value of the 1615 cm$^{-1}$ mode in p$^{in}$~-~(unpol.)$^{out}$ geometry.
Very weak Raman response is observed with the pump polarized perpendicular on the tip axis (s$^{in}$), in contrast with the case of (p$^{in}$), where strong near-field coupling gives rise to Raman enhancement. 
The spectrally broad background is due to intrinsic luminescence from the Au tip itself.  
}
\label{fig10}
\end{figure*}
 
The highly polarized TER response observed in our experiments indicates the absence of significant near-field depolarization, for the homogeneous and slender tip geometries used. 
This is required for symmetry selective probing in TERS and other nonlinear tip-enhanced processes \cite{neacsu08, lee06,gersen07,lee07} that rely on polarization selective and conserving light scattering.
In contrast, for a Ag particle-topped quartz AFM probe as used for probing the 520 cm $^{-1}$ Raman band of Si \cite{poborchii05} the observed Raman depolarization has been attributed to the wide cone angle of the tip \cite{ossikovski_prb2007}.\\

\noindent
{\bf{8. TERS of molecular adsorbates}}\\
 \noindent
 In Fig.~\ref{fig11} representative tip-enhanced Raman spectra are shown for MG on smooth Au surfaces. 
\begin{figure}[h]
\centering
\includegraphics[width=8.3cm]{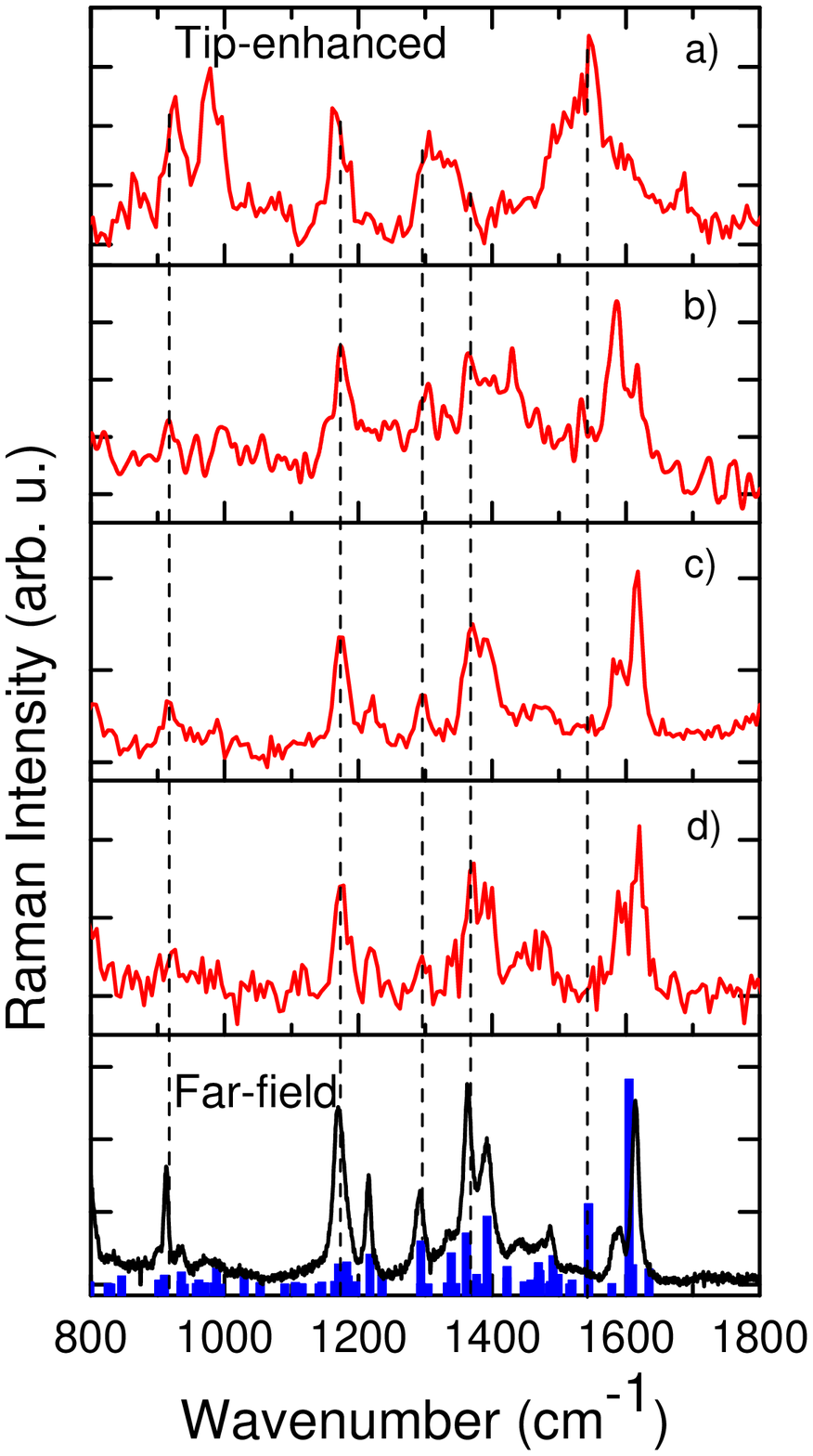} 
\caption{
Tip-enhanced  Raman spectra for $\sim 1$ ML of MG for different degrees of enhancement (a,b,c,d) in comparison with the corresponding far-field Raman spectrum (bottom graph) and DFT calculation for the mode assignment (blue bars). 
The Raman enhancement factors derived are $3\times 10^8$ (a),  $7\times 10^7$ (b), $1\times 10^7$ (c), and $1\times 10^6$ (d), respectively.  
Data are acquired for 1 s (a), 100 s (b,c), and 30 s (d), respectively. 
Spectral resolution is 25 cm$^{-1}$ for the near-field and 1 cm$^{-1}$ for the far-field spectra \cite{neacsu07b}.}
\label{fig11}
\end{figure}
They are taken for the same surface coverage of $\sim$ 1ML, but using different tips exhibiting enhancements of $3\times 10^8$ (a), $7\times 10^7$ (b), $1\times 10^7$ (c), and $1\times 10^6$ (d), respectively. 
The experimental uncertainty is estimated at a factor of 3 -- 5 for each value.
The tip-enhanced Raman spectra are reproducible for a given tip. 
But, as seen from the data, the spectral details vary from tip to tip. 
With the lateral confinement of the tip-enhancement within a $\sim$ 10 nm diameter surface region and a molecular density of $\sim 1/$nm$^2$ we estimate that the signal observed in Fig.~\ref{fig11} a)~-~d) originates from $\sim$ 100 molecules. 

In the lower panel of the figure the far-field Raman spectrum from the same sample is shown for comparison (black line). 
The far-field spectrum closely resembles that in aqueous solution \cite{lueck93} indicating that the molecules are physisorbed in isotropic orientation at the surface.  
The blue bars represent normal Raman modes of the MG anion calculated using density functional theory as implemented in Gaussian03 
as discussed in \cite{neacsu06}.
The assignment and spectral position of the calculated modes agree well with literature values \cite{lueck93, schneider95}. 
 
For moderate near-field enhancements $\leq$ 10$^7$, both spectral positions and relative intensities of the modes resemble the far-field signature, as seen in Fig.\ref{fig11} c) and d) and in accordance with other TERS results for similar enhancement \cite{domke06}. 
However, with increasing enhancement of $7\times 10^7$ (b), and most pronounced for $3\times 10^8$ (a), the vibrational modes start to look markedly different. 
While some of the modes are present in both, far-field and near-field spectra (\emph{e.g.}, 920 cm$^{-1}$, 1170 cm$^{-1}$, and 1305 cm$^{-1}$), some other are only present in the highly enhanced near-field results (\emph{e.g.}, 1365 cm$^{-1}$, 1544 cm$^{-1}$). 
The vertical dashed lines in Fig.~\ref{fig11} are added as a guide to the eye for easier comparison.

The DFT calculation allows for the identification of the spectral features in the far-field data of Fig.~\ref{fig11}. 
The majority of the intense Raman modes can be attributed to modes either localized at the phenyl ring or delocalized over the two dimethylamino phenyl groups. 
In the spectral region of  910 cm$^{-1}$ to 980 cm$^{-1}$ several vibrational modes, typically characterized by in-plane skeletal bending and/or out-of-plane C$-$H motions, are found.
The 1170 cm$^{-1}$ mode may be assigned to a methyl group rocking mode or an in-plane C$-$H bending mode of the phenyl ring. 
Around 1300 cm$^{-1}$ in-plane C$-$H deformation modes and C$-$C stretching modes of the methane group are located.  

Furthermore, the calculations show that the new spectral features seen in the highly enhanced near-field spectra in Fig.~\ref{fig11} mostly correspond to vibrational normal modes of MG. 
Among the characteristic near-field enhanced modes, \emph{e.g}, the peak at 1365 cm$^{-1}$, which is very strong in the far-field spectrum, but decreases with increasing enhancement, can be assigned to combination of the C=C stretching motions of the aromatic ring.
In contrast, the prominent peak at 1544 cm$^{-1}$ which dominates for the highest enhancement is very weak for small enhancements or in the far-field spectra. 
Here, the calculation shows a mode characterized by stretching motions combined with in-plane C$-$H bending motions of the conjugated di-methyl-amino-phenyl rings. 
The two modes at 1585 and 1615 cm$^{-1}$, which can be assigned to C=C stretching vibrations of the phenyl ring, decrease with enhancement. 

This change in both intensity and spectral signature with increasing near-field enhancement together with the vibrational analysis shows that the peaks observed may well correspond to vibrational modes of MG, whereby different selection rules must apply for the Raman spectra obtained under condition of high enhancements \cite{moskovits85}.

In the following we discuss possible physical mechanisms leading to the experimental observations.
Due to the high localization of the optical near-field, the molecules in the tip-sample gap experience a large field gradient and different Raman symmetry selection rules can come into play \cite{polubotko81, sass81,creighton88}. 
This Gradient Field Raman effect (GFR) \cite{ayars00} and the mechanism by which strong field gradients can influence the molecular Raman spectra by altering the selection rules require that the polarizability tensor ($\alpha_{ab}$) and $(dE_b/dQ)_{Q=0}$  must simultaneously be nonzero \cite{ayars00}, where E is the electric field, Q the vibrational coordinate and a,b $\in$ \{x,y,z\}.
The resulting selection rules resemble the surface selection rules \cite{moskovits82}, and give rise to GFR spectral lines which complement the Raman spectra. 
The presence of the field gradient may also lead to IR modes becoming Raman active \cite{ayars00}.
In the case of MG, it was verified that the few IR modes show no noticeable resemblance with the highly enhanced TER spectra (details in \cite{neacsu06}).
While it is difficult to quantify the contribution of GFR in general \cite{ayars00}, strong circumstantial evidence from the occurrence of normally forbidden modes, akin the observation in the TERS results suggest that the process contributes in this case. 
Similar differences between far- and near-field Raman spectra were reported previously in fiber-based SNOM experiments on Rb-doped KTiOPO$_4$ \cite{jahncke95, jahncke96}.
  
In addition, the optical field gradient can also couple to vibrations via the derivative of the quadrupole polarizability $A_{\rm ijk}$ of a mode ($\propto \partial A_{\rm ijk} / \partial\,Q\,{\bf \nabla} {\bf E}$) \cite{creighton88}.   
With $\alpha_{ab}$ and $A_{\rm ijk}$ transforming differently in terms of symmetry and their ratio being highly mode dependent this could also account for the mode selectivity observed in tip-enhanced Raman scattering.  
However, the DFT calculations of $A_{\rm ijk}$ are still deemed challenging for large molecules yet they would be desirable and can contribute to a unified description of the underlying processes, given the well characterized structural environment in tip-enhanced Raman in contrast to most SERS experiments.
It might also help to resolve the striking observation that the strength of the calculated four modes at 846, 988, 1029, and 1544 cm$^{-1}$ are overestimated by DFT calculation as compared to the far-field spectra, but coincidentally represent modes that are relatively strongly enhanced in the TER spectra. 

The pronounced spectral differences between the tip-enhanced and far-field Raman response resemble observations made in SERS, where  vibrational modes which are normally not Raman allowed had been found \cite{erdheim80, dornhaus80,moskowits80}. 
Aside from orientational effects, these spectral variations are typically interpreted to arise from conformational changes and/or transient covalent binding of the molecule at "active sites" \cite{moskovits82}. 
With that being unlikely in the tip-enhanced Raman geometry discussed here, this could indicate that purely electromagnetic mechanisms might already induce the kind of spectral selectivity observed in our TERS experiments. \\

\noindent
{\bf{9. Molecular bleaching}}\\
 \noindent
 An important question regarding the appearance of different spectral features in the near-field tip-enhanced Raman spectra is the influence of the molecular bleaching or other decomposition products \cite{maher02, wang05}. 
 \begin{figure*}[t]
  \begin{center}
   \includegraphics[width=12 cm]{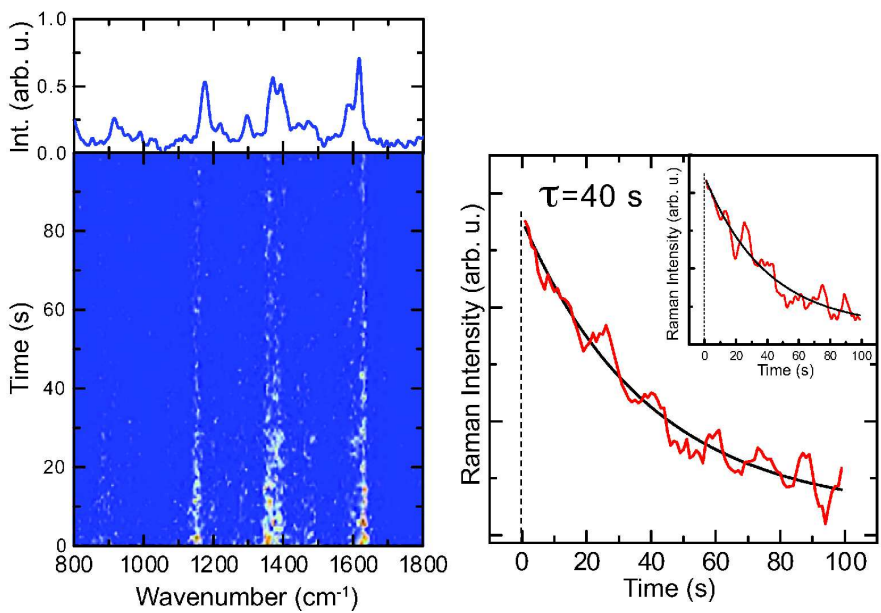} 
    \caption{
    Left: Time series of 100 consecutive near-field Raman spectra (acquired for 1 s each) for $\sim$ 1 ML MG on Au (Raman enhancement $1.3 \times 10^7$). 
    The signal decays to zero due to bleaching and no new spectral features appear from the photoreaction products.
The sum Raman spectrum (top panel) clearly resembles the far-field spectrum. 
 Right: Bleaching kinetics derived for the spectrally integrated Raman time series and the region from 1550 to 1650 cm$^{-1}$ (inset) \cite{neacsu07b}.}
\label{fig12}
  \end{center}
\end{figure*}
With the molecules under investigation exposed to the strongly localized and enhanced near-field, this leads to a sometimes rapid photo-decomposition process, especially in a resonant Raman excitation. 

To probe for the possible appearance of photoreaction products and their signature in the Raman spectra the evolution of the Raman emission is monitored in time-series experiments. 
Fig.~\ref{fig12} (left panel) shows consecutive near-field Raman spectra acquired for 1 s each for an enhancement of $1.3 \times 10^7$ with an incident laser fluence of $3 \times 10^4$ W/cm$^2$.
The molecules bleach on a time scale of $\sim 100$ s, depending on the enhancement level, and the decay and subsequent disappearance of the spectral response is uniform, \emph{i.e.}, the relative peak amplitudes are maintained. 
During the bleaching no new spectral features appear from possible photoreaction products and the signal decays with the relative peak amplitudes remaining constant. 
After complete bleaching no discernible Raman response can be observed. 
The fluctuation observed in the time series is expected given the small number of  $\sim 100$ molecules probed in the near-field enhanced region under the tip apex.  
The sum over all spectra (top graph) or the sum of any large enough subset even at later times, \emph{i.e.}, after substantial bleaching has already occurred, closely resembles the far-field response of MG and thus allows to attribute the Raman response to MG molecules.
 
The same behavior of a gradual and homogeneous disappearance of the Raman response without a relative change in peak intensity is also observed for larger enhancements, \emph{i.e.}, the case where a different mode structure is observed. 
However, the larger local field experienced by the molecules leads in general to decreasing decay time constants. 

 Fig.~\ref{fig12} (right panel) shows the decay kinetics of the spectrally integrated Raman intensity for the time series data shown on the left.
This is shown in comparison to the integral intensity of the region from 1550 to 1650 cm$^{-1}$ encompassing only the two prominent modes (inset).
Assuming an exponential decay behavior of $I/I_0 = {\rm exp}(-t/\tau)$ for the Raman intensity a decay time $\tau= 40 \pm 5\,$s is derived from the fit in both cases (solid lines). 
From the applied laser fluence of $3 \times 10^4$ W/cm$^2$ and the enhancement of the pump intensity of $\sim \sqrt{1.3 \times 10^7}$ the bleaching would be induced by a local pump fluence of $4.7 \times 10^7$ W/cm$^2$. 

\begin{figure}[t]
\centering
\includegraphics[width=8.3 cm]{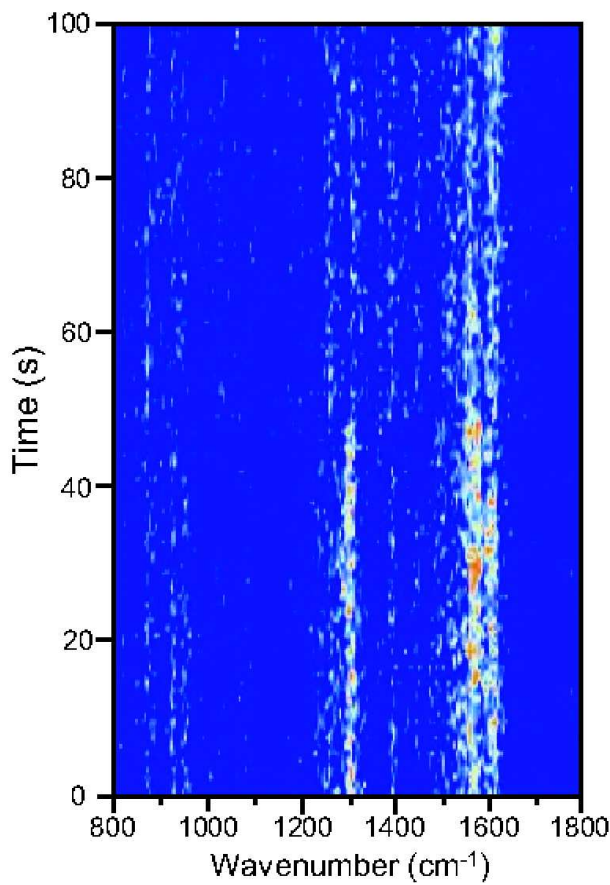} 
\caption{
Time series of 100 NF Raman spectra acquired for 1s each. 
After $\sim$50 s the Raman intensity reduces due to bleaching, but no signature of secondary products is visible.
}
\label{fig13}
\end{figure}
An extreme case of molecular bleaching is shown in Fig.~\ref{fig13}.
A time series of 100 consecutive Raman spectra from a sample with submonolayer molecular coverage, with each spectrum acquired for 1s is recorded. 
The estimated Raman enhancement factor is 9 $\times$ 10$^7$ and the spectral features deviate from the far-field Raman spectrum presented in Fig.~\ref{fig11} b, akin to the ones shown in the same figure panels a) and b) for high enhancement level (\emph{e.g.}, the mode at 1544 cm$^{-1}$). 
After an illumination time of about 50 s, the overall Raman intensity drops suddenly over the whole spectral region, and most visible for the peak at 1306 cm$^{-1}$. 
With this being a strong indication that the probed molecules undergo a bleaching process, it is interesting to note that no new spectral features emerge. 
Thus even for extreme cases as the one presented here, the molecular decomposition products do not contribute to the observed Raman signal. 
It was suggested that the molecular bleaching rate could be used for the derivation of the enhancement factor \cite{pettinger04}.
It should be noted that the bleaching rate is not a characteristic physical quantity universal for a given molecule. 
For example,  malachite green isothiocyanate~-~a sister dye of MG~-~ was found to bleach with a rate constant more than two orders of magnitude higher than the MG studied here \cite{pettinger04}, if renormalized to the same experimental conditions. 
Bleaching mechanisms can be quite diverse \cite{xie98}. 
They may include irreversible photoinduced or even multi-photon induced reactions such as rearrangements, dissociation and fragmentation, elimination or hydrogen abstraction or perhaps photooxidation with ambient oxygen via triplet states.
It can depend on, \emph{e.g.}, humidity or cleanliness of tip and sample, and is hence not a useful measure to compare experiments performed under different conditions in different laboratories. 

With the experiments carried out under ambient conditions, special care must be taken to use clean samples and tips.
It is well known \cite{moyer00} that contaminating carboniferous species and/or carbon clusters can adhere to either tip or sample and reveal their Raman signature in the tip-enhanced spectra, \emph{e.g.}, Raman bands of appreciable intensity at frequencies above 1750 cm$^{-1}$ \cite{kudelski00}.

A carboniferous Raman response (see Fig.~6 in \cite{neacsu07b}) manifests itself in characteristic spectral features from carbon clusters much different from the spectral response discussed above for monolayer or submonolayer MG.
In accordance with previous observations \cite{kudelski00,picardi,moyer00}, the carbon Raman response is comparatively large and fluctuates rapidly in an uncorrelated way. 
In contrast to the data on MG, a distinct spectral feature emerges around 2000 cm$^{-1}$ which has been assigned to, \emph{e.g.}, modes within the segments of carbon chains \cite{kudelski00}, and which is absent in the TER spectra of MG. 
This absence of the carbon Raman response can be understood since the bleaching of monolayer and submonolayer MG coverage leads to smaller molecular fragments and subsequently to a dilute surface carbon distribution and hence does not lead to extended carbon chains and aggregates which can readily form by multilayer MG decomposition at ambient temperatures.  

Besides the degradation of the analyte, another potential source of carbon contamination is the near-field probe itself. 
At room temperature and in a non-controlled atmosphere, contamination molecules from the environment could stick to the tip surface and reveal highly enhanced Raman signals.
TERS control experiments with the bare tip and clean Au samples prior to MG deposition were carried out to confirm the absence of vibrational Raman signature. 
Only intrinsic tip luminescence could be observed in that case \cite{neacsu06}. \\

\noindent
{\bf{10. TERS with single molecule sensitivity}}\\
\noindent
With the measured tip-enhanced Raman spectra presenting a signal-to-noise ratio of more than 40:1 when probing $\sim$100 molecules for 1 s accumulation time, this demonstrates the potential for even single molecule sensitivity. 
For the subsequent experiment we resort to a sample prepared with sub-monolayer surface coverage, adjusted to expect on average $<1$ molecule under the tip-confined area of $\sim 100$ nm$^2$. 

Corresponding near-field tip-enhanced Raman spectra measured in a time series with 1 s acquisition time for each spectrum are shown in Fig.~\ref{fig14} (left panel). 
Here the tip has been held at a constant distance $d$ = 0 nm above the sample and the total Raman enhancement is estimated at 5 $\times$ 10$^9$ \cite{note}.
The observed Raman signal exhibits temporal variations of relative peak amplitudes and fluctuations in spectral position.
These are characteristic signatures of probing a single emitter in terms of an individual molecule. 
Similar observations have been made before in SERS \cite{nie97,xu99,bjerneld00, dieringer07} with the fluctuations in the spectroscopic signature of a single emitter typically attributed to changes in its local environment, its structure, molecular diffusion \cite{kneipp97, futamata05}, and changes in molecular orientation \cite{wang05}. 
\begin{figure*}[t]
\centering
\includegraphics[width=17.8 cm]{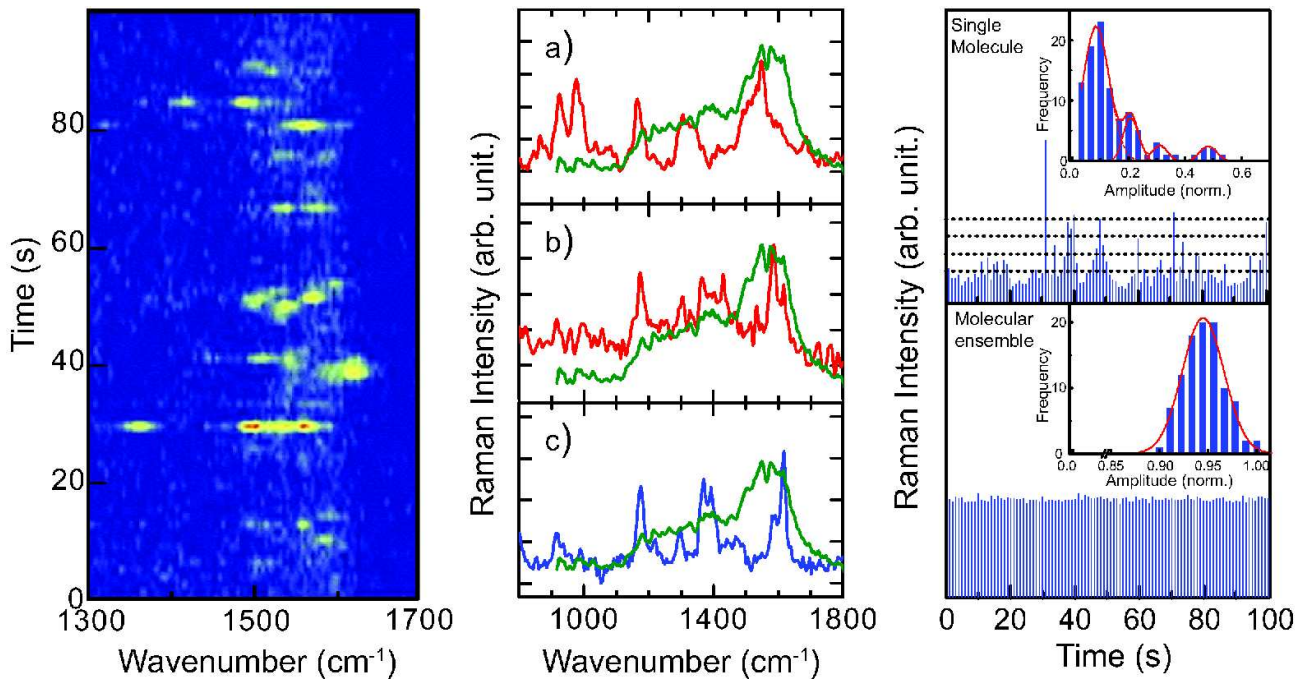} 
\caption{
Left: Time series of tip-scattered Raman spectra for a sub-monolayer MG surface coverage with Raman enhancement of 5$\times$10$^9$. 
The spectral diffusion observed is characteristic for observing single MG molecules. 
Middle: Comparison between sum spectrum of data shown in left panel (olive) and tip-enhanced Raman spectra for different degrees of enhancement (red, a) and b)) and sum spectrum of data shown in Fig.~\ref{fig12} (blue).
Right: Temporal variation of the Raman intensity of the integrated 1480 - 1630 cm$^{-1}$ of time series shown in left panel (top). 
From the corresponding histogram (inset) a discretization of Raman intensities can be seen with 170-230 counts$\cdot\rm molecule^{-1} \cdot s^{-1}$ (dashed line increment). 
For comparison, temporal variation of the Raman intensity of the integrated 1480 - 1630 cm$^{-1}$ of big molecular ensemble (far-field) is shown (bottom) together with corresponding histogram (inset) \cite{neacsu06, neacsu07b}.}
\label{fig14}
\end{figure*}
With MG only physisorbed, it has to be considered in particular that the molecules can diffuse in and out of the apex-confined probe region while experiencing different degrees of enhancement. 
The diffusion dynamics can be facilitated by the thin water layer present on the sample surface under ambient conditions. 

In the time series in Fig.~\ref{fig14} the apparent bleaching rate seems reduced compared to what is expected from the analysis of the ensemble bleaching discussed above.
This is a result of the low surface coverage where new molecules directly neighboring the tip-sample gap can diffuse into the near-field enhanced region. 
However, the signal vanishes  rapidly after 100 s due to the eventual depletion after the molecules surrounding the near-apex area have bleached. 

A different mode structure of MG is observed compared to the far-field response~-~highly visible for the 1500 -- 1600 cm$^{-1}$ spectral region, akin to the results for strong enhancements shown above. 
Fig.~\ref{fig14} (middle panel) displays the sum spectrum (olive) for the single molecule response time series (left) in comparison with ensemble spectra of $\sim$ 100 molecules probed for one second (red) showing different degrees of enhancement: 3$\times$10$^8$ in a) and 7$\times$10$^7$ in b).
Panel c) presents the sum spectrum (blue) of the data shown in Fig.~\ref{fig12} a) which, as for a moderate enhancement, closely resembles the far-field spectrum. 
The resemblance of the Raman spectrum of a molecular ensemble with the sum spectrum over the whole time series offers strong evidence of probing single intact MG molecules \cite{weiss01}.
Note that the sum over the times series shall therefore resemble an ensemble spectrum for large enhancement, rather than the far-field Raman spectrum.
In assessing the resemblance of the spectral characteristics it has to be considered that at low coverage the molecules have more degrees of freedom to dynamically change orientation and they can diffuse.
Given the rather weak response, the signal detected can only be expected to emerge from the region of largest enhancement, and with the diffusing molecules probing the spatial variation of the enhancement under the tip this corresponds to an extreme case of inhomogeneous broadening. 
Therefore, while individual spectral features at positions in accordance with the strongly tip-enhanced near-field response are observed in the time series, the sum spectrum no longer exhibits clearly resolved lines. 
This interpretation is further sustained considering, \emph{e.g}, the improved resemblance of the peak in the 1550 -- 1600 cm$^{-1}$ region of the single molecule sum spectrum with the sum of the two near-field spectra (in a and b) of different enhancement. 

Further insight is obtained by studying the statistical behavior of the single molecule Raman response \cite{kneipp97}.
Fig.~\ref{fig14} (right panel top graph) displays the integrated 1430 cm$^{-1}$ to 1650 cm$^{-1}$ spectral intensity for the data in the left panel. 
The signal intensities cluster with intervals of 170-230 counts$\cdot \rm s^{-1}$, as already evident from visually inspecting the time-series of integrated intensities (dashed lines), and this  manifests itself in the corresponding histogram in an asymmetric distribution with discrete peaks (inset). 
This behavior is qualitatively reproducible for experiments with the same surface coverage and it can  be interpreted as the Raman emission from $n=0$ (noise peak), 1, 2, and 3 molecules being probed under the tip, as suggested for similar findings in SERS \cite{kneipp97, willets07}. 
This assignment is corroborated from experiments with different surface coverages: for lower coverages only the $n=0$ and 1 peaks remain and with increasing coverage the distribution converges to a narrow random Gaussian distribution which is observed from a large molecular ensemble, as seen in the lower panel, where 100 consecutive far-field spectra were acquired for 100 s each and the signal is integrated over the same spectral range as in the case of the single molecule experiment.
The details of the histogram, however, depend on the binning procedure especially for a small data set as has been shown to be insufficient as the sole argument for single molecule observation \cite{leru06}. 

The optical trapping and alignment of MG under the tip must also be considered as a possible source of the observed surface diffusion and intensity fluctuations \cite{friedrich95}. 
However, for MG together with our particular experimental conditions, this can not explain the discretization of Raman peak intensities in the single molecule response, as detailed elsewhere \cite{neacsu06}.  

Our single molecule TER results are similar to other recent experimental findings \cite{zhang07} where brilliant cresyl blue molecules adsorbed on planar Au were probed. 
For enhancements of $\sim$ 10$^7$ similar to our intermediate values, temporal fluctuations in both intensity and mode frequency were observed, albeit with no significant spectral differences between the far-field and near-field response as seen here for high enhancements.\\

\noindent
{\bf{11. Raman imaging of nanoscrystals: Near-Field Crystallographic Symmetry}}\\
\noindent
TERS can provide symmetry selective information. 
This potential, however, has yet been largely unexplored. 
Here we show that the characteristic symmetry properties of the tip scattering Raman response in terms of independent control of polarization and $k$ vector for both incident and Raman scattered light provides the necessary degree of freedom to determine the crystallographic orientation and -domains in nanostructures.
This is due to the fundamental interaction with lattice phonons of the Raman process, together with the symmetry properties and selection rules of the tip-enhanced scattering  that allows for identifying crystallographic axes without requiring atomic resolution. 
With Raman scattering being less invasive than electron or x-ray techniques and applicable {\it{in situ}}, this approach will fill a much needed gap in the characterization of nanostructured materials with increasing complexity \cite{chen_l2007,zhu_em2007}.

Transmission Electron Microscopy, capable of providing atomic resolution \cite{wang_am2003}, requires samples thin enough to be transparent for electrons, extensive sample preparation, and due to vacuum conditions, makes {\it{in situ}} experiments difficult \cite{zhang_2001}.
Likewise, X-ray microscopy is capable of characterizing nanostructures with atomic resolution\cite{zuo_sci2003}, but requires a monochromatic brilliant synchrotron radiation source and radiation beam damage remains a concern \cite{chapman_josaa2006}.
Here, the comparable simplicity of TERS from an instrumentation perspective makes it highly attractive providing complementary information and even avoiding some of the disadvantages of the existing techniques. 

In Raman spectroscopy, the specific phonon modes probed depend on the chosen experimental geometry, in terms of the incident and detected polarization as well as the propagation direction of light \cite{hendra_a1970, didomenico_pr1968}.
These phonon modes allow determination of the crystallographic orientation of a sample. 
This has been shown in far-field Raman in, \emph{e.g.} the study of $90^\circ$ domain switching in bulk BaTiO$_3$ \cite{li_jap1990} or the observation of ferroelastic domains in LaNiO$_4$ \cite{nakamura_jpsj1990}. 
However, in Raman microscopy, in the commonly used confocal epi-illumination and detection geometry, this reduces the available degrees of freedom, and thus losing the general capability of probing the symmetry specific Raman tensor elements.
In extending the use of the Raman selection rules to a side-illuminated TERS geometry, these degrees of freedom can be regained and even further refined by taking into account the tip geometry as discussed in the following.

The intensity of the Raman scattered light from a medium is given by: $I_s  \propto \vert \vec{e_s}\cdot \bar{R} \cdot \vec{e_i} \vert ^{2}$ where $\vec{e_i}$ and $\vec{e_s}$ are the polarization of the incident and scattered light, respectively and the Raman tensor $\bar{R}$  is the derivative of the susceptibility tensor \cite{gardiner_1989}. 
As an example, for a Raman-active phonon mode of tetragonal BaTiO$_3$, $\bar{R}$ is given by:
\begin{equation*} 
\mathbf{A_1 (\vec{\xi})} = 
\left(
 \begin{array}{ccc}
a & 0 & 0\\
0&a&0\\
0&0&b
\end{array}
\right)
\end{equation*}
where $\vec{\xi}$ denotes the polarization direction of the mode (for polar modes). The symmetry of a given mode, in this case $A_1$, is determined from group theory and may contain multiple component phonon modes of different frequency \cite{drago_1992}. 
Thus, when the polarization conditions, determined from the susceptibility derivative, are satisfied for a given symmetry mode, the Raman shift due to the phonons belonging to that mode can be observed.

In addition, one can selectively isolate specific phonon modes within a symmetry mode. 
For polar modes, the phonons will separate into Transverse Optical (TO) and Longitudinal Optical (LO) components \cite{arguello_pr1969}, which, being distinct in frequency, can be spectrally resolved.
This is not accounted for by the Raman tensor methods described above, requiring further refinement of the selection rules.

For a given geometry, the wavevector $\vec{q}$ of the propagating phonon can be determined by conservation of momentum from the wavevectors of the incident and scattered light. 
Based on the relative orientation between $\vec{q}$ and $\vec{\xi}$, one can selectively excite the LO mode for $\vec{q} \parallel \vec{\xi}$, or the TO mode for $\vec{k} \perp \vec{\xi}$.
Thus,  the observation of either a TO or an LO mode provides the additional information about the orientation of the crystallographic axes.

Furthermore, drawing on the nanoscopic apex of a plasmonic tip for preferential enhancement of incident light polarized along the tip axis allows us to exploit the unique symmetry selection rules associated with the tip \cite{neacsu05}. 
This will enhance modes whose polarizations coincide with this axis. 
Therefore we expect maximum enhancement of modes for which both allowed incident and scattered polarization directions are oriented parallel with the tip axis.
Modes for which either the incident or scattered polarization coincide with the enhancement axis may also be observed, albeit with a lower intensity.

The potential for applying the symmetry properties of the Raman selection rules in a tip-enhanced geometry has been emphasized previously \cite{ossikovski_prb2007,lee_jrs2007, nguyen_sd2007,saito_apl2006}. 
However, no general treatment has yet been discussed.
Here we will briefly summarize the essential elements for tip-enhanced Raman spectroscopy of nanocrystals, with further details to be published elsewhere \cite{berweger}.
\begin{figure*}[t]
\centering
\includegraphics[width=12cm]{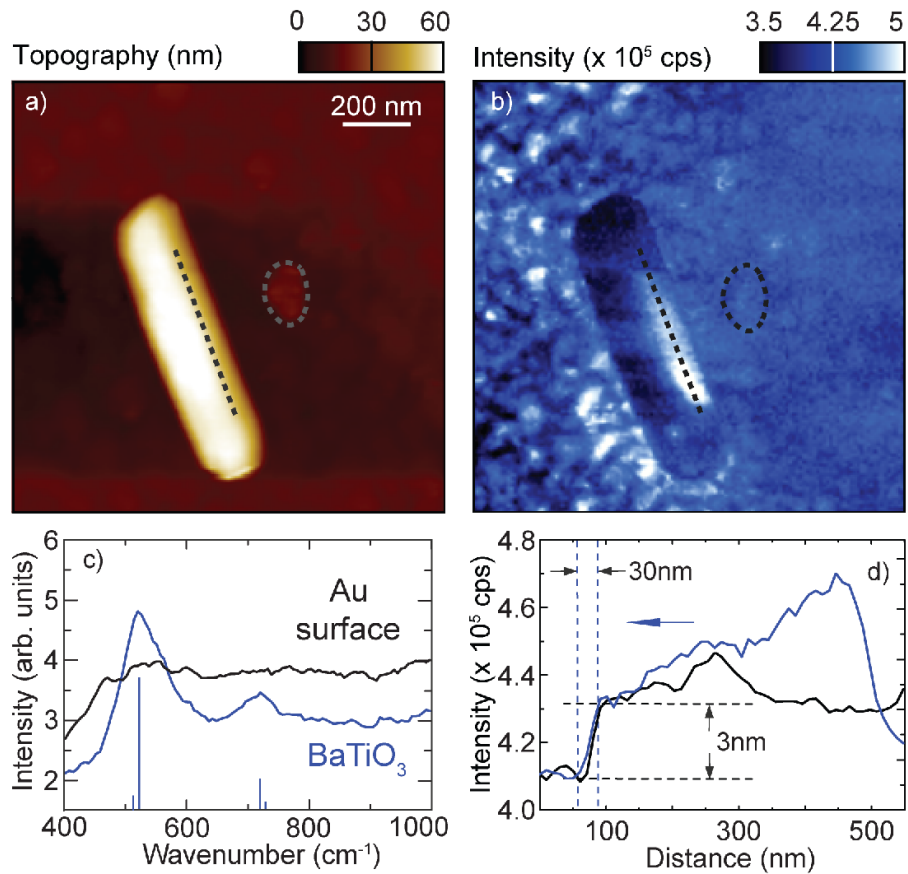} 
\caption{
a) Shear-force topography of a crystalline BaTiO$_3$ nanorod on a Au substrate (1.28 $\times$ 1.28 $\mu$m$^2$).
b) Spectrally integrated TERS signal from the same surface region, showing strong optical contrast on top of the nanorod, as well as above highly localized substrate regions.
The enclosed region (dashed line) shows low signal enhancement due to dielectric contrast from sample contamination.
c) Raman spectra acquired on a BaTiO$_3$ rod (blue) and on Au substrate (black).
The peaks are identified (blue vertical lines) as E$_1$ TO mode at $510$ cm$^{-1}$, a second order peak at $520$ cm$^{-1}$ and a combination of the A$_1$ LO and E$_1$ LO modes at $727$ cm$^{-1}$ and $715$ cm$^{-1}$, respectively.
A cross section of the region of high enhancement (blue) and corresponding topography (blue) on the rod is shown in panel d), taken as an average over 3 adjacent pixels along the straight dashed line in panels a) and b ).
The strong rise in optical signal (left) is correlated with a $\sim$ 3 nm height variation, unlike the sharp decrease in optical response (right) which has no topographic correspondent; this seems to indicate the presence of a crystal defect.}
\label{fig15}
\end{figure*}

In order to demonstrate the conceptual capability of TERS on nanocrystals, we studied individual single crystal BaTiO$_3$ nonorods in the tetragonal phase \cite{mau_jacs2003}. 
Perovskite BaTiO$_3$, a displacive ferroelectric at room temperature, has long been the focus of intense study due to interest in these properties \cite{blinc_sb2007}.
Although ferroelectric domain formation and crystallographic orientation on the macroscale are well understood, complementary studies on the nanoscale are desired.

TERS acquired from a BaTiO$_3$ rod is shown in Fig.~\ref{fig15}. 
The BaTiO$_3$ nanorods were placed on a Au-coated quartz substrate.
Panel a) shows the shear-force topography of a rod, with the corresponding spectrally integrated TERS signal acquired simultaneously in panel b).
The incident polarization is parallel to the tip axis and the detected signal is unpolarized with an acquisition time of $10$ ms for each pixel.

On the Au substrate itself, topographically localized  regions (down to $30$ nm) can be identified, of intense luminescence enhancement due to the plasmonic tip-sample coupling \cite{pettinger07}.
These regions are correlated with small nanometer-scale Au surface features.	
In contrast, the enclosed area in the figure shows a clear topography but a comparatively weak enhancement of the optical signal.
This suggests a surface contamination resulting in dielectric contrast but little or no plasmonic coupling.

Panel a) shows TERS spectra acquired on top of a nanorod (blue) and on the Au substrate (black).
The peaks at $520$ cm$^{-1}$ and $720$ cm$^{-1}$ identify the material as BaTiO$_3$ and can be assigned to a higher order peak and a combination of the A$_1$ LO and E$_1$ LO modes, respectively \cite{didomenico_pr1968}. 
The spectral positions and estimated relative intensities of the peaks are shown by the vertical lines underneath the curve.
The spectrally broad luminescence background observed originates form the Au tip itself.

Panel d) shows cross section profiles taken from panels a) and b) (black and blue, respectively).
Spatial resolution of $30$ nm can be estimated and is correlated with tip apex.
The cross section is taken from the region of strong Raman enhancement on the rod, as indicated by the straight dashed lines in a) and b).
The signal shown is the average over 3 consecutive pixels disposed orthogonal to the cross-sectional direction.
A clear correlation between a $3$ nm rise in the topography with a strong increase in the optical signal can be seen.
This indicates a distinct region of the rod responsible for the increased signal.
Due to distinct topographic features of BaTiO$_3$ previously observed at grain boundaries \cite{gheno_fe2006}, we attribute the signal observed in this region to different crystallographic subdomains in the rod \cite{buscaglia_prb2006}.

Although the study of nanocrystalline samples opens up a wide range of potential applications for TERS, some fundamental aspects are not yet fully understood.
Recent far-field studies of wurtzite CdS nanorods indicate a possible depolarization effect in dielectric nanostructures, leading to a breaking of the Raman tensor selection rules \cite{fan08}. 
Furthermore, it has been shown that the presence of a sharp edge within the near-field of a photoemitter can affect the polarization of the emitted light \cite{moerland08}, although the resulting effect on Raman scattering is yet unclear.
In addition, the large field gradient near the tip can fundamentally alter the selection rules, making previously silent modes visible \cite{ayars00}. 
Although this may render IR and other modes Raman-active \cite{ayars01}, making mode assignment more difficult, this would shed further insight into the fundamental material properties.\\

\noindent
{\bf{12. Outlook}}\\
\noindent
Reproducibility can be enhanced performing TERS under controlled experimental environmental conditions. 
Performing experiments under \emph{e.g.}, Ultra High Vacuum (UHV) conditions offers variable sample temperature and combination with other UHV techniques for surface analysis \cite{steidtner07, berndt91}. 
Tip enhanced Raman spectroscopy may emerge as an important analytical tool for chemical and structural identification on the nanoscale.
It offers chemical specificity, nanometer spatial resolution, single molecule sensitivity and symmetry selectivity. 
However, with both sensitivity and spatial resolution critically dependent on the well defined geometry and related optical properties of the tip, reproducibility has remained an issue in TERS.
Furthermore, for direct far-field illumination conditions, it is difficult to \emph{a priori} distinguish the near-field response from the unspecific far-field imaging artifacts (\emph{vide supra}).  
\begin{figure}[t]
\centering
\includegraphics[width=8.3 cm]{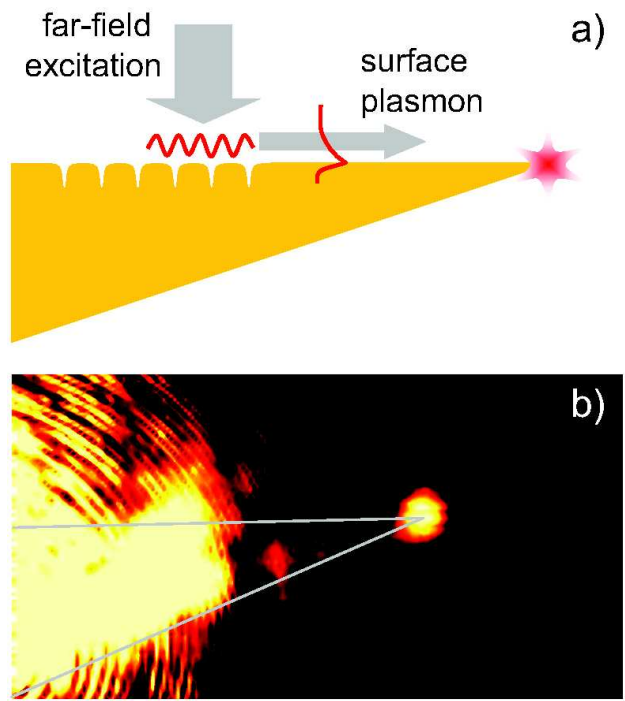} 
\caption{
a) Principle of the nonlocal excitation of the tip apex. 
Far-field radiation excites SPP on the grating, which propagate along the shaft towards the tip apex, where they are reradiated into the far-field.
b) Microscope image recorded for illumination of the tip-grid demonstrating the efficient nonlocal excitation of the tip apex via illumination of the grating \cite{ropers07}.}
\label{fig16}
\end{figure}

Future developments in tip design and fabrication may prove critical. 
We have recently demonstrated a novel way to generate a nanoconfined light emitter on a nanoscopic probe tip obtained by grating-coupling of surface plasmon polaritons on the tip-shaft \cite{ropers07}. 
The adiabatic field concentration of the propagating SPP, determined by the boundary conditions imposed by the tapered shape of the tip, offers an intrinsic nano-focusing effect and thus gives rise to confined light emission only from the apex region, as theoretically predicted  \cite{stockman04}. 

In this experiment linear gratings are written onto the shaft of Au tips by focused ion beam milling, $\sim$ 10 $\mu$m away from the apex, as schematically shown in Fig.~\ref{fig16} a).
Upon grating illumination with a broadband light source (150 nm spectral bandwidth of Ti:sapphire oscillator), SPP are excited and launched towards the apex \cite{raether, ebbesen98}, where they are re-radiated, as shown in Fig.~\ref{fig16} b) (details discussed in \cite{ropers07}). 

Spatially separating the excitation from the apex itself, this approach is particularly promising as it avoids the otherwise omnipresent far-field background present for direct apex illumination. 
In addition, it would provide the spatial resolution needed for near-field optical techniques including \emph{s}-SNOM and TERS. \\

\noindent
{\bf{13. Summary}}\\
\noindent
A systematic understanding both experimentally and theoretically of the fundamental processes responsible for field enhancement, spectral tip plasmonic response and tip-sample coupling has allowed for reaching Raman enhancement factors as high as 10$^9$ leading to the ultimate sensitivity limit in analytical spectroscopy -- the single molecule. 
With lateral resolution solely determined by the tip apex radius, nanometer spatial resolution can be obtained. 
Criteria have been discussed for experimental TERS implementation and distinction of near-field signature from far-field imaging artifacts.
The combination of the inherent sensitivity and spatial resolution of TERS with the Raman selection rules and the unique symmetry of the scanning tip makes possible the spatially resolved vibrational mapping on the nanoscale. 
Its implementation for determining the orientation and domains in crystallographic nanostructures has been proposed.
Future developments in tip design and more efficient illumination and detection geometries and scanning probe implementation will allow for TERS to become a powerful nano-spectroscopic analysis tool.\\

\noindent
{\bf{Acknowledgment.}} The authors would like to thank Nicolas Behr, Jens Dreyer, Thomas Elsaesser, Christoph Lienau and Claus Ropers  for valuable discussions and support, and Stanislaus Wong for providing the BaTiO$_3$ sample. 
Funding by the Deutsche Forschungsgemeinschaft through SFB 658 ("Elementary Processes in Molecular Switches at Surface"), the National Science Foundation (NSF CAREER grant CHE 0748226) is greatly acknowledged. 
S. B. thanks the National Science Foundation for the IGERT Fellowship.

 \bibliographystyle{bmc_article}  
\bibliography{ters2}

\begin{thebibliography}{100}
\bibitem{note2}
 [In addition to the unique symmetry properties, the local surface and nonlocal
  longitudinal bulk polarizations contributions to the nonlinear polarization
  and their directional and polarization selection rules are directly
  distinguishable here due to the geometry of the tip as a partial asymmetric
  ($\infty m m$) nanostructure with the mirror symmetry broken along the
  axis.].
\bibitem{note1}
 [Note that for the calculations that refer to the field at the tip apex, the
  field is calculated at 0.125 nm below the apex to avoid numerical artifacts
  due to finite grid size].
\bibitem{note}
 [Here d = 0 nm is defined as corresponding to a 20-30 \% decrease in the
  shear-force amplitude.].
\end{thebibliography}

\end{document}